\documentclass[11pt]{article}
\usepackage[T1]{fontenc}
\usepackage[utf8]{inputenc}
\usepackage{lmodern}
\usepackage[margin=1in,headheight=14pt]{geometry}
\usepackage{microtype}
\usepackage{needspace}
\usepackage{amsmath,amssymb,amsthm,mathtools}
\usepackage{booktabs,array,tabularx,enumitem}
\usepackage{algorithm}
\usepackage{algpseudocode}
\usepackage[dvipsnames]{xcolor}
\usepackage{tikz}
\usepackage{hyperref}
\hypersetup{colorlinks=true,linkcolor=MidnightBlue,citecolor=MidnightBlue,
 urlcolor=MidnightBlue,pdftitle={Tight Bounds for Differentially Private Cut Release},
 pdfauthor={Chenghua Liu and Boning Meng}}
\usepackage[capitalize,noabbrev]{cleveref}
\makeatletter
\def\theHALG@line{\thealgorithm.\arabic{ALG@line}}
\makeatother
\newtheorem{theorem}{Theorem}[section]
\newtheorem{lemma}[theorem]{Lemma}
\newtheorem{proposition}[theorem]{Proposition}

\newtheorem{claim}[theorem]{Claim}
\theoremstyle{definition}

\numberwithin{equation}{section}
\newcommand{\R}{\mathbb R}
\newcommand{\E}{\mathbb E}
\newcommand{\eps}{\varepsilon}
\newcommand{\ip}[2]{\langle #1,#2\rangle}
\newcommand{\norm}[1]{\lVert #1\rVert}
\newcommand{\one}{\mathbf 1}
\DeclareMathOperator{\diag}{diag}
\DeclareMathOperator{\tr}{tr}
\DeclareMathOperator{\SDP}{SDP}
\DeclareMathOperator{\Lap}{Lap}
\newcommand{\KL}{D_{\mathrm{KL}}}

\DeclareMathOperator{\poly}{poly}
\DeclareMathOperator*{\argmax}{arg\,max}
\DeclareMathOperator*{\argmin}{arg\,min}
\definecolor{pinkfoldpink}{HTML}{B74B7D}
\newcommand{\pinkfoldmark}{%
  \raisebox{-0.15em}{%
    \begin{tikzpicture}[x=0.0375em,y=-0.0375em]
      \path[use as bounding box] (0,0) rectangle (24,24);
      \draw[pinkfoldpink,line width=0.06375em,line join=miter,line cap=butt]
        (12,2) -- (20.660,7) -- (20.660,17) -- (12,22) --
        (3.340,17) -- (3.340,7) -- cycle
        (12,12) -- (20.660,7)
        (12,12) -- (3.340,7)
        (12,12) -- (12,22);
    \end{tikzpicture}%
  }%
}
\allowdisplaybreaks[2]
\setlist{itemsep=2pt,topsep=5pt}
\title{Tight Bounds for Differentially Private Cut Release}
\author{%
Chenghua Liu\textsuperscript{1} and Boning Meng\textsuperscript{2}\\[0.5em]
\small \textsuperscript{1}Institute of Software, Chinese Academy of Sciences, Beijing, China\\
\small \textsuperscript{2}University of Regensburg, Regensburg, Germany\\[0.3em]
\small \texttt{liuch.russell@gmail.com}, \texttt{mengboning2013@gmail.com}}
\date{}
\begin{document}
\maketitle
\begin{abstract}
We give a polynomial-time algorithm for releasing a nonnegative weighted
synthetic graph under $(\eps,\delta)$-edge differential privacy.
Given an $n$-vertex nonnegative weighted graph $G=(V,w_G)$, the output
$H=(V,w_H)$ satisfies, with probability at least $1-\beta$,
simultaneously for every $S\subseteq V$,
$|w_G(S,V\setminus S)-w_H(S,V\setminus S)|
\le\gamma w_G(S,V\setminus S)+O\bigl((n\log n)/(\eps\gamma)\bigr)$,
provided that $\gamma,\eps,\delta,\beta$ are at least inverse polynomial
in $n$. For constant $\gamma,\eps$ and $\delta=\beta=n^{-10}$, this
improves the recently obtained additive error bound of $O(n\log^7 n)$ to $O(n\log n)$.
We also prove a matching $\Omega(n\log n)$ lower bound.
Our algorithm combines degree decomposition with private multiplicative
weights, using a degree-sensitive semidefinite oracle to generate update
proposals from single Gaussian samples. Each proposal is validated by a
noisy check of its exact entropy decrease, and a joint R\'enyi analysis
of sampling and validation avoids additional logarithmic losses.
\end{abstract}
\setlength{\abovedisplayskip}{10pt plus 3pt minus 6pt}
\setlength{\belowdisplayskip}{10pt plus 3pt minus 6pt}
 \section{Introduction}\label{sec:introduction}
Graphs that record relationships between individuals, such as social,
financial, or communication networks, often contain sensitive
information. Differential privacy~\cite{DMNS06} provides a rigorous
framework for publishing statistics of such graphs; we use edge-level
privacy, which protects the presence and weight of individual
relationships while treating the vertex set as public. A central problem
in this setting is to release a synthetic graph that approximately
preserves every cut. Cut values underlie cut maximization, balanced
partitioning, and clustering, and the released graph can be
postprocessed freely, at no additional privacy cost.

Formally, we consider a graph $G=(V,w_G)$ on a public vertex set $V=[n]$
with nonnegative edge weights $w_G\in\R_{\ge0}^{\binom n2}$. For disjoint
sets $S,T\subseteq V$, we write $w_G(S,T)$ for the total weight of edges
between them, so the cut defined by $S$ has weight $w_G(S,V\setminus S)$.
Two inputs $G,G'$ are adjacent if $\norm{w_G-w_{G'}}_1\le1$,
so privacy protects every change of total edge weight at most one,
whether concentrated on one edge or spread over many. The weights may be
arbitrary nonnegative reals; neither their support nor a bound on their
magnitude is public. Given a relative error $\gamma>0$, we seek a
polynomial-time differentially private mechanism that releases a
nonnegative weighted graph $H=(V,w_H)$ satisfying, simultaneously for every cut
$S$,
\begin{equation}\label{intro:guarantee}
 |w_G(S,V\setminus S)-w_H(S,V\setminus S)|
 \le\gamma w_G(S,V\setminus S)+B,
\end{equation}
with additive error $B$ as small as possible. The relative term makes
the guarantee informative for sparse graphs: with purely additive error
($\gamma=0$), an error of order $n^{3/2}$, up to logarithmic factors, is
unavoidable in the worst case over graphs with $\Theta(n^2)$
edges~\cite{EKKL20,LUZ24}. This error can exceed every cut value of a
sparse graph.

Table~\ref{tab:prior} summarizes the known bounds. In the purely
additive setting, independent noise on every edge weight gives additive
error $\widetilde O(n^{3/2}/\eps)$, where $\widetilde O$ hides
logarithmic factors, including the dependence on $\delta$. Following
earlier work by Gupta, Roth, and Ullman~\cite{GRU12} and
Eli\'{a}\v{s}, Kapralov, Kulkarni, and Lee~\cite{EKKL20}, Liu, Upadhyay,
and Zou~\cite{LUZ24} obtained $\widetilde O(\sqrt{mn}/\eps)$ additive
error for graphs with $m$ weighted edges,
which is tight for dense graphs up to logarithmic factors. With
relative error, Blocki, Blum, Datta, and
Sheffet~\cite{blocki2012johnson} obtained nearly linear additive error
by applying the exponential mechanism to the exponentially large family
of sparse graphs; no polynomial-time implementation is known. Aamand et al.~\cite{aamand2025breaking} gave the first
polynomial-time mechanism below the $n^{3/2}$ scale, with additive error
$n^{5/4+o(1)}$ for constant $\gamma$ and $\eps$, using a private
expander decomposition, and Fan et al.~\cite{fan2026private} reduced the
additive error to $n^{13/12+o(1)}$ for unweighted inputs through
degree-sensitive private spectral primitives.

Most recently, Eli\'{a}\v{s} et al.~\cite{EliasEtAl26} gave the first
polynomial-time algorithm with nearly linear additive error,
$B=O(\eps^{-1}\gamma^{-1}n\log^7(n/(\gamma\eps\delta\beta)))$ for
arbitrary nonnegative weighted graphs. They also prove a lower bound of $\Omega((n/\eps)\min\{\sqrt
n,1/\gamma\})$ for $0<\gamma\le1/4$, $\delta\le\eps/100$, and
$\beta\le1/40$. For fixed $\gamma$ and $\eps$ and inverse-polynomial
$\delta$ and $\beta$, the optimal additive error was therefore known to
lie between $\Omega(n)$ and $O(n\log^7n)$. We show that it is
$\Theta(n\log n)$: a single logarithmic factor is both sufficient, in
polynomial time, and necessary.

\begin{table}[t]
\centering
\small
\begin{tabular}{@{}llll@{}}
\toprule
Result & Privacy & Running time & Guarantee\\
\midrule
\cite{EKKL20} & approximate & polynomial
 & $w_G(S,V\setminus S)\pm\widetilde O\bigl(\sqrt{\norm{w_G}_1n}\bigr)$\\
\cite{LUZ24} & approximate & polynomial
 & $w_G(S,V\setminus S)\pm\widetilde O(\sqrt{mn})$\\
\cite{blocki2012johnson} & pure & exponential$^*$
 & $(1\pm\gamma)w_G(S,V\setminus S)\pm n\poly(\log n)$\\
\cite{aamand2025breaking} & approximate & polynomial
 & $(1\pm\gamma)w_G(S,V\setminus S)\pm n^{5/4+o(1)}$\\
\cite{fan2026private}$^\dagger$ & approximate & polynomial
 & $(1\pm\gamma)w_G(S,V\setminus S)\pm n^{13/12+o(1)}$\\
\cite{EliasEtAl26} & approximate & polynomial
 & $(1\pm\gamma)w_G(S,V\setminus S)\pm O(n\log^7n)$\\
Theorem~\ref{thm:main} & approximate & polynomial
 & $(1\pm\gamma)w_G(S,V\setminus S)\pm O(n\log n)$\\
\midrule
Lower bound~\cite{EliasEtAl26} & approximate & ---
 & $B=\Omega(n)$\\
Theorem~\ref{thm:intro-lower} & approximate & ---
 & $B=\Omega(n\log n)$\\
\bottomrule
\end{tabular}
\caption{Error bounds for private all-cut release on $n$-vertex graphs,
with $m=\norm{w_G}_0$. We fix constant $\eps,\gamma$ and
$\delta=\beta=n^{-c}$ for constant $c\ge2$; $\widetilde O$ hides
polylogarithmic factors. $^*$Information-theoretic exponential mechanism.
$^\dagger$Unweighted inputs.}
\label{tab:prior}
\end{table}

\subsection{Our results}

\begin{theorem}[Weighted cut release]\label{thm:main}
Fix a constant $c\ge1$. Let $n\ge2$,
$n^{-c}\le\gamma\le1/4$, $n^{-c}\le\eps\le1$, and
$n^{-c}\le\delta,\beta<1/10$. There is an
$(\eps,\delta)$-differentially private algorithm that, given any
nonnegative weighted graph $G=(V,w_G)$ on $V=[n]$, outputs a nonnegative
weighted graph $H=(V,w_H)$ satisfying, with probability at least
$1-\beta$, simultaneously for all $S\subseteq V$,
\begin{equation}\label{intro:full-bound}
 |w_G(S,V\setminus S)-w_H(S,V\setminus S)|
 \le\gamma w_G(S,V\setminus S)+O\Bigl(\frac{n\log n}{\eps\gamma}\Bigr).
\end{equation}
The implied constant depends only on $c$. For rational inputs, the
running time is polynomial in $n$ and the input encoding length.
\end{theorem}

The inverse-polynomial assumptions allow the additive error to be written
as $O((n/(\eps\gamma))\log n)$. Without these assumptions, the analysis
gives the more general bound in~\eqref{global:full-bound}, which makes
the dependence on $\gamma,\eps,\delta,\beta$ explicit.

\begin{theorem}[Lower bound]\label{thm:intro-lower}
Fix a constant $\kappa\in(0,1/2)$. For sufficiently large $n$, let
$n^{-1/2+\kappa}\le\gamma\le1/4$, $0<\eps\le1$,
$0\le\delta\le\eps/n^2$, and $0\le\beta\le1/4$. Any
$(\eps,\delta)$-differentially private mechanism that, on every
nonnegative weighted input graph $G$ on $[n]$, outputs a nonnegative weighted graph
$H$ satisfying~\eqref{intro:guarantee} simultaneously for all cuts with
probability at least $1-\beta$ must have
\begin{equation}\label{intro:lower}
 B=\Omega\Bigl(\frac{n\log n}{\eps\gamma}\Bigr).
\end{equation}
The implied constant depends only on $\kappa$.
\end{theorem}

Together, Theorems~\ref{thm:main} and~\ref{thm:intro-lower} determine
the optimal additive error $\Theta(n\log n/(\eps\gamma))$ under their
common parameter assumptions. In particular, the optimal error is
$\Theta(n\log n)$ for constant $\gamma,\eps$ and $\delta=\beta=n^{-10}$.
Section~\ref{sec:lower} gives a more general lower bound.

\subsection{Technical overview}\label{sec:overview}

The logarithm in our bound reflects the cost of making many noisy
decisions accurate simultaneously while keeping the total privacy loss
small. We account for these costs, together with the logarithm in the
initial entropy potential, through the parameter
\[
 \sigma=\Theta\Bigl(\frac1\eps\log\frac{n}{\gamma\eps\delta\beta}\Bigr).
\]
The guiding principle of the design is to pay this logarithm only once:
neither the number of degree layers nor the number of trials should
introduce another logarithmic factor. In the parameter range of
Theorem~\ref{thm:main}, our mechanism achieves a total additive error of
$O(n\sigma/\gamma)$. We first recall the framework we build on and then
describe the three ideas that achieve this.

\paragraph{Private multiplicative weights with a semidefinite oracle.}
Our mechanism is a private multiplicative-weights search~\cite{HR10}. It
maintains a public candidate graph $h$, repeatedly looks for a cut on
which $h$ is inaccurate, and then scales the edges of $h$ across that
cut up or down, depending on whether $h$ underestimates or overestimates
the cut. Progress is measured by the relative entropy between the private
graph and $h$, which decreases with every good update. The most violated
cut cannot be used directly: finding it is NP-hard, and small input
changes can change the maximizing cut abruptly, making its direct
release incompatible with privacy. As in~\cite{EKKL20,fan2026private,EliasEtAl26}, we instead
solve a semidefinite relaxation of the cut problem regularized by
$\lambda\log\det X$. Its optimizer $X$ is a covariance matrix, and a
Gaussian vector $z\sim N(0,X)$ assigns each pair $uv$ a nonnegative
score $(z_u-z_v)^2/4$, giving a random update direction. The log-determinant
term makes $X$ insensitive to small changes of the private graph, so the
sample $z$ can be released at a cost of $O(\alpha/\lambda^2)$ in
R\'enyi divergence of order $\alpha$, provided $\alpha/\lambda$ is
sufficiently small (Lemma~\ref{oracle:lem-rdp}).

The price of this stability is accuracy. The regularization blurs the
cut that the oracle finds, and Lemma~\ref{oracle:thm-harmonic} shows
that the resulting error is a sum over the vertices, in which a vertex of
effective degree $d$ in the oracle contributes about
$\lambda^2/(\lambda+\gamma d)$.
A vertex whose degree is large compared with $\lambda/\gamma$ thus
contributes little, whereas a vertex of small degree contributes about
$\lambda$. The regularization parameter therefore
governs a trade-off: a large $\lambda$ makes each sample cheap in
privacy, and a small $\lambda$ makes the cut accurate.

\paragraph{Degree layers.}
Since the error of a vertex depends on its degree, no single value of
$\lambda$ suits all vertices. We therefore decompose the graph by degree
and release each part with its own regularization parameter. A private
peeling procedure repeatedly removes the vertices whose remaining degree
falls below a threshold, where the thresholds $D_0<D_1<\cdots$ double
from stage to stage. The vertices removed at stage $j$ form the layer
$L_j$, and each edge is assigned to the lower layer of its endpoints.
Layer $j$ is released with $\lambda_j=\sqrt{\sigma D_j}$, a choice that
balances privacy and accuracy within the layer. On the accuracy side,
every vertex of a layer $j\ge1$ has degree at least $D_j/4$, so it
contributes $O(\lambda_j^2/(\gamma D_j))=O(\sigma/\gamma)$ to the error;
in layer zero, $\lambda_0=\Theta(\sigma/\gamma)$ is small enough by
itself. The corresponding privacy balance is explained below.

One difficulty remains: a vertex of a higher layer also belongs to the
vertex sets of the lower layers, where
its degree may be small and its error contribution large. We release the
layers from the top down and add to the oracle a public prior built from
the already released higher layers, with geometrically decreasing
weights. The prior supplies these vertices with enough degree that a
vertex $k-j$ layers above layer $j$ contributes only a $(2/3)^{k-j}$
fraction of $O(\sigma/\gamma)$, so the total error remains
$O(n\sigma/\gamma)$.

The oracle's error also contains a term
proportional to the degree discrepancy between the candidate and the true
layer. We release noisy degrees of each layer and confine the search for
layer $j$ to a public convex confidence set $\mathcal K_j$ of graphs whose
degrees, mass, and minimum degrees are consistent with this information.
With high probability, $\mathcal K_j$ contains the true layer, which
therefore remains a valid target for the relative-entropy potential, and
the total degree discrepancy over all layers is only $O(n\sigma)$.
Finally, since the oracle needs bounded edge weights and no bound on the
input weights is public, a thresholded Laplace mechanism first releases
the part of every edge weight above a cap $U=\Theta(\sigma/\gamma)$,
which costs relative error at most $\gamma/2$.

\paragraph{One sample per trial, checked before use.}
Within a layer, the search proceeds in trials. A trial draws a single
Gaussian sample from the oracle and turns the resulting edge scores
into a bounded multiplicative update of the candidate. A single sample
is a noisy direction: it makes progress only with constant probability
(Lemma~\ref{piece:one-proposal}), and averaging many samples to obtain a
reliable direction would multiply the privacy cost of a trial by their
number. Instead, we check each sample before using it. The decrease of
the potential caused by the update has an exact closed form in terms of
the sample and the private graph, and it depends linearly on the latter.
The trial adds Laplace noise to this decrease and applies the update
only if the noisy value is large; otherwise the candidate stays unchanged.
On the high-probability event that all validation noises are bounded,
every accepted update makes definite progress. Rejected proposals leave
the potential unchanged. Since the potential of a layer of mass $M$ is initially at most
$O(M\log n)$, only a bounded number of updates can be accepted, while a
candidate with a large error is improved with constant probability in
every trial. An accurate candidate therefore appears within a bounded
number of trials, and the exponential mechanism selects one at the end
(Lemma~\ref{piece:release}).

\paragraph{Privacy accounting.}
A trial reveals both the Gaussian sample and the noisy validation. The
sample is private by the stability of the oracle. For the validation,
the point is that the updates are gentle: before projection, their
multiplicative factors lie in $[1/2,2]$, and the absolute values of their
logarithms are at most the step size times the sampled edge scores.
A change of the private graph therefore moves the
validated quantity by at most the step size times a weighted average of
the sampled edge scores. The validation noise has scale proportional to
the step size times $\lambda$. We control the exponential moments of
this random shift and analyze the sample and its validation jointly.
For an input change of total edge weight $s\le1$, a trial costs
$O(\alpha s^2/\lambda^2)$ in R\'enyi divergence of order $\alpha$,
whether or not its proposal is accepted
(Lemma~\ref{privacy:trial}). Rejected proposals thus need no separate
accounting.

A layer with threshold $D_j$ uses $O(\eps D_j)$ trials, each costing
$O(\alpha s_j^2/(\sigma D_j))$, where $s_j$ is the total edge-weight
change in that layer. Its total trial cost is therefore
$O(\alpha\eps s_j^2/\sigma)$. Conditional on the preprocessing
transcript, each edge belongs to exactly one layer, so a unit change of
the input satisfies $\sum_j s_j\le1$ and hence $\sum_j s_j^2\le1$.
Thus composition incurs no extra factor for the number of layers.
Composing all trials in R\'enyi divergence and
converting to approximate differential privacy once at the end proves
privacy (Proposition~\ref{global:privacy}). The accuracy guarantees add
up in the same way: every vertex contributes $O(\sigma/\gamma)$, the
remaining error terms of the layers grow geometrically with $\lambda_j$
and give the second term of~\eqref{global:full-bound},
and the relative errors add up because the discounted priors have total
weight at most three times that of the released graph
(Section~\ref{sec:global}).

\paragraph{The lower bound.}
The lower bound shows that one logarithm is the unavoidable price of
hiding which edges are present. Consider random bipartite graphs with
$\bar n=\lfloor n/2\rfloor$ vertices on each side, in which every left
vertex has $d$ neighbors and every edge has weight $W$. We decode the
private output to a graph in the same family. If we delete a random edge,
its missing endpoint is uniform among about $\bar n$ positions, so the
decoder, run on the graph without this edge, recovers the endpoint with
probability $O(d/\bar n)$. Restoring the edge is a change of $W$ units,
and group privacy raises this probability by at most a factor
$e^{\eps W}$, plus a $\delta$-dependent additive term. Taking $W$ to be
a sufficiently small constant multiple of
$\eps^{-1}\min\{\log(\bar n/d),\log(\eps/\delta)\}$ therefore leaves
most edges unrecovered; the second logarithm is infinite when $\delta=0$.
Accurate cut values would rule this out: random signs on the right
vertices, with the left signs chosen to maximize the discrepancy, give a
random cut with expected discrepancy of order $W\bar n\sqrt d$, while
the expected combined cut value is $W\bar nd$. Choosing
$d=\Theta(\min\{n,\gamma^{-2}\})$, with a sufficiently small constant,
prevents the relative term $\gamma W\bar nd$ from absorbing this
discrepancy. The additive error must therefore be
$\Omega(W\bar n\sqrt d)$, yielding~\eqref{lower:general} and
hence~\eqref{intro:lower}. 

\subsection{Comparison with Eli\'{a}\v{s} et al.}

Like~\cite{EliasEtAl26}, we release the heavy part of each edge weight
by a thresholded Laplace mechanism, find violated cuts with a
log-determinant-regularized semidefinite program accessed through
Gaussian samples~\cite{EKKL20,fan2026private}, and check sampled
proposals with noise before using them. The difference lies in how
progress is organized. Their mechanism follows the PREM
framework~\cite{PREM25}: with $L=\Theta(\log(n/(\gamma\eps\delta\beta)))$,
it runs $O(L)$ rounds of $O(L/\gamma^2)$ multiplicative-weights steps,
and each step uses an aggregate witness built from many fractional
witnesses, each estimated by averaging $O(L)$ Gaussian samples. Their
analysis~\cite[Section~5.2]{EliasEtAl26} incurs two factors of $L$ from
the rounds, through error accumulation and privacy composition, one
from composition over the steps, and two from success-charged
composition within an aggregation. The high-probability bound on the privacy loss of the Gaussian samples
and the certificate term of order $\lambda n\log n$ each contribute
another factor, giving additive error $O(nL^7/(\gamma\eps))$.

We replace the rounds with a single degree decomposition whose layers
add up in both error and privacy without a factor for their number.
Each trial uses one Gaussian sample and tests
a noisy estimate of its entropy decrease before projection, ensuring
with high probability that every accepted update makes definite
progress without aggregation. Since a trial has the same R\'enyi
cost whether or not it is accepted, all trials are composed in R\'enyi
divergence and converted to approximate differential privacy once.
Together with our degree-sensitive certificate, which has no $n\log n$
term, this ensures that the only logarithm in our bound is the one in
the noise scale $\sigma$.

\paragraph{Organization.}
Section~\ref{sec:preliminaries} gives preliminaries,
Section~\ref{sec:algorithm} the mechanism, Section~\ref{sec:analysis}
its analysis and the proof of Theorem~\ref{thm:main}, and
Section~\ref{sec:lower} the proof of Theorem~\ref{thm:intro-lower} and
its consequences. Appendix~\ref{sec:implementation} describes a
finite-precision implementation.
 \section{Preliminaries}\label{sec:preliminaries}
We fix notation and recall the semidefinite relaxation, the privacy
primitives, and the relative entropy used as our potential. All
logarithms are natural. Unless stated otherwise, asymptotic notation
hides absolute constants, and the letters $C,c$ denote positive absolute
constants whose values may change from one occurrence to the next; the
algorithmic constants are fixed in the order specified below. Products
$x\odot y$, powers, and logarithms of vectors are taken coordinatewise,
and $\ip{x}{y}$ denotes the usual inner product.

\paragraph{Graphs and cuts.}
We identify a weighted graph on the public vertex set $V$ with its edge
vector in $\R^{\binom V2}$ and write $w=w_G$ for the input weights in
the algorithm and its analysis. Edge vectors on subsets of $V$ are
extended by zero when needed. For any edge vector $x$, we use
$x(S)$ as shorthand for $\sum_{u\in S,\,v\in V\setminus S}x_{uv}$;
in particular, $w(S)=w_G(S,V\setminus S)$.
For a signed edge vector $w$, such as a
difference of two graphs, let $A_w$ be its symmetric adjacency matrix
with zero diagonal, $\deg(w)$ its degree vector, and
$L_w\coloneqq\diag(\deg(w))-A_w$ its Laplacian. We write $\deg_w(v)$ for
the coordinate of $\deg(w)$ at $v$. For $w\ge0$, the mass of $w$ is
$\|w\|_1=\frac12\sum_v\deg_w(v)$.

For a symmetric matrix $X$, define the fractional cut vector $q(X)$ by
\[
 q_{uv}(X)\coloneqq\frac14(e_u-e_v)^\top X(e_u-e_v),
 \qquad\text{so that}\qquad
 \ip{w}{q(X)}=\frac14\tr(L_wX)
\]
for every signed edge vector $w$. If $X\succeq0$ and $X_{vv}\le1$ for
every $v$, then $q_e(X)\in[0,1]$ for every pair $e$. For a cut vector
$s\in\{-1,1\}^V$, the vector $q(ss^\top)$ is the incidence vector of the
corresponding cut. We use the semidefinite relaxation of Goemans and
Williamson~\cite{GoemansWilliamson95},
\[
 \SDP(w)\coloneqq\max\bigl\{\ip{w}{q(X)}:X\succeq0,\ \diag X=\one\bigr\}.
\]
It is nonnegative (take $X=\one\one^\top$), dominates every cut value
(take $X=ss^\top$), and is $1$-Lipschitz in $\ell_1$, that is,
$|\SDP(w)-\SDP(w')|\le\|w-w'\|_1$, because $q(X)\in[0,1]^{\binom V2}$
for every feasible $X$.

\paragraph{Differential privacy.}
Inputs $w,w'\ge0$ are adjacent if $\|w-w'\|_1\le1$; they may differ in
many coordinates. A mechanism $\mathcal M$ is
$(\eps,\delta)$-differentially private if, for every ordered pair
$(w,w')$ of adjacent inputs and every measurable event $A$,
\[
 \Pr[\mathcal M(w)\in A]\le e^{\eps}\Pr[\mathcal M(w')\in A]+\delta.
\]
The case $\delta=0$ is called pure privacy. We use postprocessing and
adaptive composition in their standard forms~\cite{DMNS06,DR14}. Group
privacy is needed only for the lower bound of Section~\ref{sec:lower}:
if two inputs $w,w'$ are joined by $k$ unit changes, then
\[
 \Pr[\mathcal M(w)\in A]\le e^{k\eps}\Pr[\mathcal M(w')\in A]
 +\delta\sum_{i=0}^{k-1}e^{i\eps}.
\]

The mechanism of Section~\ref{sec:algorithm} draws on three standard
purely private primitives. The Laplace distribution $\Lap(\varsigma)$
with scale $\varsigma>0$ has density
$e^{-|x|/\varsigma}/(2\varsigma)$, and adding independent $\Lap(\Delta/\eps)$
noise to the coordinates of a vector query of $\ell_1$ sensitivity
$\Delta$ is pure $\eps$-differentially private; we use it for the
heavy-edge and degree releases of Section~\ref{sec:decomposition}. To
choose among a public set of $N$ candidates, the exponential
mechanism~\cite{MT07} samples each candidate $x$ with probability
proportional to $\exp\bigl(-\eps\,\mathrm{score}(x)/(2\Delta)\bigr)$,
where $\Delta$ is the sensitivity of the score. It is pure
$\eps$-differentially private and, with probability at least $1-\beta$,
returns a candidate whose score is within $2\Delta\log(N/\beta)/\eps$
of the minimum; each layer selects its output in this way.

The third primitive, the one-positive AboveThreshold
algorithm~\cite{DR14}, performs the degree comparisons of the peeling
step. We state it for a generic privacy parameter $\eps'>0$: it uses
threshold zero, threshold noise $\Lap(2/\eps')$, and independent query
noise $\Lap(4/\eps')$. It answers a sequence of possibly adaptive
queries and stops at the first positive comparison. The density-shift
proof of its privacy bounds the likelihood ratio of its transcript by
$e^{\eps'\Delta}$ whenever every query changes by at most $\Delta$, for
any real $\Delta\ge0$. Fractional values of $\Delta$ arise because a
unit change of the input may change many degrees, each by less than one.

The trials within each layer are not of this pure form: they release
Gaussian samples, and we account for the privacy of the layer releases
through R\'enyi divergence. For $\alpha>1$, the
R\'enyi divergence of order $\alpha$ is
\[
 D_\alpha(P\Vert Q)\coloneqq\frac{1}{\alpha-1}
 \log\E_{x\sim P}\Bigl[\Bigl(\frac{dP}{dQ}(x)\Bigr)^{\alpha-1}\Bigr].
\]
A mechanism is $(\alpha,\rho)$-RDP if this divergence between its output
distributions is at most $\rho$ for every ordered pair of adjacent
inputs; it is then $(\rho+\log(1/\delta)/(\alpha-1),\delta)$-differentially
private for every $\delta\in(0,1)$~\cite{Mironov17}. We apply this
conversion only once, to all layer releases together
(Proposition~\ref{global:privacy}).

\paragraph{Relative entropy.}
For nonnegative edge vectors $f,h$, define the unnormalized relative
entropy by
\[
 \KL(f\Vert h)\coloneqq\sum_e\Bigl[f_e\log\frac{f_e}{h_e}-f_e+h_e\Bigr].
\]
We allow $f$ and $h$ to have different total masses and use the usual
conventions at zero. Each summand is nonnegative, so
$\KL(f\Vert h)\ge0$.
 \section{The algorithm}\label{sec:algorithm}
We now specify the mechanism outlined in Section~\ref{sec:overview};
Section~\ref{sec:analysis} analyzes it. The mechanism first releases the
part of every edge weight above a public cap, peels the remaining capped
graph into degree layers, and builds a public confidence set for each
layer (Section~\ref{sec:decomposition}). It then releases the layers one
at a time, in decreasing order, by the entropy search of
Section~\ref{sec:piece}.

All releases draw on a common privacy and failure budget. We set
$\eps_0=\eps/10$ and $\beta_0=\beta/10$. Each preprocessing release and
the final selection use privacy budget $\eps_0$; the trials and the
numerical implementation use the remaining allowance. Each preprocessing
accuracy event has failure probability at most $\beta_0$, and each layer
uses $\zeta=\beta_0/(J+1)$, where $J$ is the peeling
horizon~\eqref{decomp:eq-horizon}. The calibration parameter, the relative slack,
the cap, the degree thresholds, and the regularization parameters are
\begin{equation}\label{global:eq-parameters}
 \begin{aligned}
 \sigma&\coloneqq\frac{C_\sigma}{\eps}\log\frac{n}{\gamma\eps\delta\beta},
 &\qquad a&\coloneqq\frac{\gamma}{40},
 &\qquad U&\coloneqq\frac{C_U\sigma}{\gamma},\\
 D_j&\coloneqq\frac{2^jC_D\sigma}{\gamma^2},
 &\qquad \lambda_j&\coloneqq\sqrt{\sigma D_j}.
 \end{aligned}
\end{equation}
We choose the absolute constants in the order $C_\sigma,C_U,C_D$, with
$C_D$ sufficiently large that $\lambda_j\ge5U$ for every $j$ and the
noise bounds below hold.

Each of these parameters has a distinct role. The calibration parameter $\sigma$
carries the single logarithmic factor of the final bound, $a$ is the
relative slack of each layer, and $U$ is the cap above which weights are
released directly. The condition $\lambda_j\ge5U$ keeps the edge weights
seen by the oracle below $\lambda_j$, as Lemma~\ref{oracle:thm-harmonic}
requires. Finally, the choice $\lambda_j^2=\sigma D_j$ balances the
R\'enyi cost of a trial, which scales as $1/\lambda_j^2$
(Lemma~\ref{privacy:trial}), against the harmonic error term of the
oracle, which carries a factor $\lambda_j^2$ (Section~\ref{sec:global}).

\subsection{Degree decomposition}\label{sec:decomposition}
The mechanism begins by separating the heavy part of every edge, so that
all later steps see bounded weights. We keep the capped weights
$b_e=\min\{w_e,U\}$ private and release the excess above the cap by a
thresholded Laplace mechanism,
\begin{equation}\label{decomp:eq-heavy}
 Y_e\coloneqq\bigl((w_e-U)_++Z_e-t_{\mathrm h}\bigr)_+,
 \qquad
 t_{\mathrm h}\coloneqq\frac{1}{\eps_0}\log\frac{2\binom n2}{\beta_0},
\end{equation}
where the noise variables $Z_e\sim\Lap(1/\eps_0)$ are independent. The
excess map is an $\ell_1$ contraction, so this release is pure
$\eps_0$-DP. With probability at least $1-\beta_0$, all
$|Z_e|\le t_{\mathrm h}$. Then $Y_e$ underestimates the excess by at
most $2t_{\mathrm h}$ and vanishes when the excess is zero. Since
$U\ge4t_{\mathrm h}/\gamma$, every cut satisfies
$|Y(S)+b(S)-w(S)|\le(\gamma/2)w(S)$.

We next partition the vertices by their degrees in $b$. The peeling uses
the public horizon, query bound, and noise threshold
\begin{equation}\label{decomp:eq-horizon}
 J\coloneqq\max\Bigl\{0,\Bigl\lceil\log_2\frac{2(n-1)U}{D_0}\Bigr\rceil\Bigr\},
 \qquad Q_{\mathrm c}\coloneqq n(n+J+1),
 \qquad \tau_{\mathrm c}\coloneqq\frac{C}{\eps_0}
 \log\frac{C(n+Q_{\mathrm c})}{\beta_0}.
\end{equation}
Thus $D_J\ge2(n-1)U$, twice the largest possible degree in $b$;
$Q_{\mathrm c}$ bounds the number of threshold queries, and
$\tau_{\mathrm c}$ bounds the error of all threshold comparisons with
high probability. Our parameter choices ensure
$D_0\ge16\tau_{\mathrm c}$ and $U\ge\tau_{\mathrm c}$.

The comparisons themselves are made by AboveThreshold. We give each
vertex an independent, persistent one-positive AboveThreshold instance
(Section~\ref{sec:preliminaries}) with privacy $\eps_0/2$, threshold
zero, threshold-noise scale $4/\eps_0$, and query-noise scale
$8/\eps_0$. Initially, $\mathcal W=V$. At stage $j=0,\ldots,J$, we scan
the active set $\mathcal W$ in a public order, querying
$D_j-\deg_{b[\mathcal W]}(v)$ for each vertex $v$. A
positive answer removes $v$ permanently and restarts the scan. A
complete scan without removal ends the stage; the instances persist
across stages. Let $L_j$ contain the vertices removed at stage $j$, with
any remaining vertices assigned to $L_J$. Put $V_j=\bigcup_{k\ge j}L_k$,
the active set at the start of stage $j$, and $n_j=|L_j|$. We assign
each edge to the minimum layer of its endpoints and write
$b=\sum_jb_j$, so that $b_j$ is a graph on $V_j$. Thus, $b_j$ consists
of the edges of $b[V_j]$ with at least one endpoint in $L_j$.

The layers determine what degree information is released. We release
the concatenated degree vectors of the $b_j$ by adding independent
$\Lap(2/\eps_0)$ noise, then clamp each coordinate to $[0,(n-1)U]$,
obtaining $y^{(j)}$. With $m_j=\binom{|V_j|}2$, the degree radius and
the mass bound of layer $j$ are
\[
 R_j\coloneqq\frac{C_r}{\eps_0}
 \Bigl(|V_j|+\log\frac{C(J+1)}{\beta_0}\Bigr),
 \qquad M_j\coloneqq\min\{2n_jD_j,\,m_jU\},
\]
and the confidence set of layer $j$ consists of the graphs $h$ on $V_j$
that are consistent with the released degrees and with the peeling,
\begin{equation}\label{decomp:eq-confidence}
 \begin{split}
 \mathcal K_j\coloneqq\Bigl\{0\le h\le U:{}
 &\|\deg(h)-y^{(j)}\|_1\le R_j,\ \|h\|_1\le M_j,\\
 &\deg_h(v)\ge\frac{D_j}{4}\text{ for all }v\in L_j\text{ if }j\ge1\Bigr\}.
 \end{split}
\end{equation}
The mass and lower-degree constraints come directly from the peeling
guarantees in the following lemma, which collects the properties of the
decomposition used in the analysis.

\begin{lemma}[Degree decomposition]\label{decomp:interface}
The layer and degree releases are pure $2\eps_0$-DP. For a fixed
preprocessing transcript, adjacent inputs satisfy
\begin{equation}\label{decomp:eq-partition}
 \sum_js_j\le1,\qquad\textup{where }s_j\coloneqq\|b_j-b'_j\|_1.
\end{equation}
The construction makes at most $Q_{\mathrm c}$ threshold queries. With
probability at least $1-2\beta_0$, every $b_j\in\mathcal K_j$, and every
$h\in\mathcal K_j$ satisfies $\|\deg(h)-\deg(b_j)\|_1\le2R_j$. Moreover,
\begin{equation}\label{decomp:eq-total-degree-error}
 \sum_jR_j\le\frac{C}{\eps_0}
 \Bigl[n(J+1)+(J+1)\log\frac{C(J+1)}{\beta_0}\Bigr]
 =O(n\sigma).
\end{equation}
\end{lemma}
\begin{proof}
We begin with the privacy of peeling. Fix a transcript and put
$\Delta_v=\sum_{e\ni v}|b_e-b'_e|$. The transcript fixes the active
sets and queries, and each query of $v$ changes by at most $\Delta_v$.
The AboveThreshold density-shift proof therefore bounds the likelihood
ratio of the transcript at vertex $v$ by $e^{\eps_0\Delta_v/2}$,
including when $\Delta_v<1$. The instances use independent randomness,
so multiplying these bounds and using $\sum_v\Delta_v=2\|b-b'\|_1$
proves privacy. Each scan makes
at most $n$ queries, and there are at most $n$ removals and $J+1$
completed scans. Conditional on the partition, the concatenated degree
query has sensitivity two: each edge contributes to two coordinates of
one layer. Clamping is postprocessing. The same edge partition
gives~\eqref{decomp:eq-partition}.

For accuracy, except with probability $\beta_0$, removals have current
degree at most $D_j+\tau_{\mathrm c}$ and rejections at least
$D_j-\tau_{\mathrm c}$. Since $D_J\ge2(n-1)U$, all vertices are removed
by the last stage. The final rejection scan at stage $j-1$ has active
set $V_j$, and all of the degree of a vertex $v\in L_j$ in $b[V_j]$
belongs to $b_j$. Charging each edge to its first removed endpoint gives
\[
 \deg_{b_j}(v)\ge\frac{D_j}{4}\text{ for all }v\in L_j\text{ if }j\ge1,
 \qquad
 \|b_j\|_1\le2n_jD_j\text{ for all }j\ge0.
\]
Laplace $\ell_1$ tail bounds give, simultaneously,
$\|y^{(j)}-\deg(b_j)\|_1\le R_j$, except with probability $\beta_0$.
Clamping cannot increase this error. Together with the peeling bounds,
this proves $b_j\in\mathcal K_j$; the degree-discrepancy bound follows
by the triangle inequality. Summing $R_j$ gives
\eqref{decomp:eq-total-degree-error}, using $J=O(\log n)$ and
$\log(n/\beta)=O(\eps\sigma)$.
\end{proof}

Each confidence set also supplies the starting point of the entropy
search in its layer. We omit empty layers, and output zero for a layer
when $\mathcal K_j$ is empty or $m_j=0$. Otherwise, we solve a separate
LP for each edge coordinate, maximizing that coordinate over
$\mathcal K_j$. Let $h_j^0\in\mathcal K_j$ be the average of the
resulting optimizers. Since the solution for
coordinate $e$ has $e$th coordinate at least $h_e$ for every
$h\in\mathcal K_j$, and every $h\in\mathcal K_j$ has mass at most $M_j$,
every $h\in\mathcal K_j$ satisfies
\[
 h^0_{j,e}\ge\frac{h_e}{m_j}\text{ for every pair }e,
 \qquad
 \KL(h\Vert h_j^0)\le M_j(\log m_j+1)\le M_j\log(2n^2).
\]
We remove the coordinates with $h^0_{j,e}=0$, which vanish on all of
$\mathcal K_j$; if none remain, we output zero for the layer.

It remains to fix the order in which the layers are released and the
error each may incur. We process the remaining layers in decreasing
order of $j$. Layer $j$ uses a prior $P_j$ formed from the public
outputs $h_k$ of the higher layers and a public error allowance $A_j$,
\begin{equation}\label{global:eq-allowance}
 \begin{aligned}
 P_j&\coloneqq\sum_{k>j}(3/4)^{k-j}h_k,
 &\qquad A_j&\coloneqq C_A\Bigl(\frac{\sigma\omega_j}{\gamma}+R_j\Bigr),\\
 \omega_0&\coloneqq n,
 &\qquad \omega_j&\coloneqq n_j+\sum_{k>j}(2/3)^{k-j}n_k
 \quad\text{for }j\ge1.
 \end{aligned}
\end{equation}
We choose $C_A$ after the preceding constants. The prior compensates
for the vertices of $V_j$ that lie in higher layers: such a vertex in
$L_k$ may have small degree in $b_j$, but it has degree at least
$D_k/4$ in $h_k$, which the prior supplies to the oracle. The effective
vertex counts satisfy $\sum_j\omega_j\le4n$, and
$\sum_jP_j\le3\sum_jh_j$ coordinatewise. Section~\ref{sec:global} shows
that $A_j$ suffices for the oracle error. All parameters except the
prior are fixed by preprocessing.

\subsection{Releasing a degree layer}\label{sec:piece}
We now release a single block, dropping the layer index. The layer
mechanism takes as input a public nonempty compact convex set
$\mathcal K$ of nonnegative graphs of mass at most $M$, a public
initializer $h_0\in\mathcal K$ positive on the relative support of
$\mathcal K$ and satisfying
\begin{equation}\label{piece:initial-potential}
 \KL(h\Vert h_0)\le M\log(2n^2)\qquad\text{for all }h\in\mathcal K,
\end{equation}
and public parameters $P\ge0$ and $A,\lambda>0$; in the full mechanism,
these are $\mathcal K_j$, $M_j$, $h_j^0$, $P_j$, $A_j$, and
$\lambda_j$. The accuracy proof assumes that the private input $b$
belongs to $\mathcal K$; privacy does not require this event.

The search is guided by the log-determinant oracle. For a signed edge
vector $E$ and a nonnegative edge vector $T$ on the current vertex set,
we define
\begin{equation}\label{oracle:eq-optimizer}
 X(E,T)\coloneqq\argmax_{Z\succ0,\ Z_{ii}\le1}
 \Bigl\{-\frac14\tr(A_EZ)-a\ip{T}{q(Z)}+\lambda\log\det Z\Bigr\}.
\end{equation}
For a candidate $h$, we put $E=h-b$ and $T=h+b+P$, and set
\begin{equation}\label{piece:covariance}
 X_\theta\coloneqq X(\theta E,T),\qquad
 V(h)\coloneqq\max_{\theta\in\{-1,1\}}\SDP(\theta E-aT).
\end{equation}
Since $\SDP$ dominates every cut value, every cut satisfies
$|h(S)-b(S)|\le aT(S)+V(h)$, so a candidate with small $V(h)$ is
accurate. The oracle relates $V(h)$ to the covariance score
\[
 Q_\theta(h)\coloneqq\theta\ip{E}{q(X_\theta)}
 -\frac{a}{4}\ip{T}{q(X_\theta)^2}.
\]
Lemma~\ref{oracle:thm-harmonic} will give the certificate
\begin{equation}\label{piece:oracle-interface}
 \SDP(\theta E-aT)\le4Q_\theta(h)+A
 \qquad\text{for all }h\in\mathcal K\text{ and }\theta\in\{-1,1\}.
\end{equation}
A large cut discrepancy therefore yields a covariance with a large score.

To act on such a covariance, we turn a Gaussian sample into a bounded
multiplicative update and validate its exact entropy decrease. We set
$\tau=a/2880$ and, for $z\sim N(0,X_\theta)$, put
\begin{equation}\label{piece:actual-score}
 g\coloneqq q(zz^\top),\qquad
 r_e\coloneqq\frac{1+2\tau g_e}{1+\tau g_e},\qquad
 \Delta_\theta\coloneqq\ip{h}{1-r^{-\theta}}-\theta\ip{b}{\log r}.
\end{equation}
The vector $g$ has mean $q(X_\theta)$, and $1\le r_e\le2$. When
$b\in\mathcal K$, the decrease in $\KL(b\Vert h)$ before projecting
$u=h\odot r^{-\theta}$ onto $\mathcal K$ is exactly $\Delta_\theta$,
and Lemma~\ref{piece:one-proposal} shows that
$\Delta_\theta\ge\tau Q_\theta(h)$ with constant probability.
The oracle score $Q_\theta(h)$ thus links the cut violation $V(h)$ to
potential progress, while a noisy check of $\Delta_\theta$ determines
whether the proposal is accepted. For fixed $h,r,\theta$, the input $b$
enters $\Delta_\theta$ only through $-\theta\ip{b}{\log r}$. Since
$0\le\log r_e\le\tau g_e$, the validation noise below has scale
proportional to $\tau$.

The number of trials and the acceptance threshold are calibrated so
that, with high probability, every accepted update makes definite
progress. For a failure
allocation $\zeta\in(0,1/4)$, we take a sufficiently large absolute
$C_N$ and set
\begin{equation}\label{piece:horizon}
 N\coloneqq\Bigl\lceil C_N\Bigl[\frac{M\log(2n^2)}{\tau A}
 +\log\frac{6}{\zeta}\Bigr]\Bigr\rceil.
\end{equation}
The first term is the initial potential
bound~\eqref{piece:initial-potential} divided by the minimum entropy
decrease $\tau A$ of an accepted update. We fix an absolute
$c_{\mathrm v}\ge2$ and define
\begin{equation}\label{piece:threshold}
 \ell_N\coloneqq\log\frac{6(N+1)}{\zeta},\qquad
 \Gamma\coloneqq\tau(A+c_{\mathrm v}\lambda\ell_N).
\end{equation}
When every validation noise has magnitude at most
$c_{\mathrm v}\tau\lambda\ell_N$, which a union bound over the trials
provides with high probability, the acceptance threshold $2\Gamma$
ensures that every accepted update has $\Delta_t\ge\Gamma\ge\tau A$.

\begin{algorithm}[htbp]
\caption{\textsc{ReleaseLayer}$(b)$}\label{piece:algorithm}
\begin{algorithmic}[1]
\Require Private block $b$; public layer data $\mathcal K$, $h_0$, $P$,
 $M$, $A$, $\lambda$; parameters $a$, $\tau$, $\eps_0$, $\zeta$
\Ensure A graph in $\mathcal K$
\State Compute $N$, $\ell_N$, and $\Gamma$ by~\eqref{piece:horizon}
 and~\eqref{piece:threshold}
\State Store the initial candidate $h_0$
\For{$t=0,\ldots,N-1$}
 \State Draw $\theta_t$ uniformly from $\{-1,1\}$; compute
 $X_{\theta_t}$ at $h_t$ by~\eqref{piece:covariance}
 \State Draw $z_t\sim N(0,X_{\theta_t})$; form $r_t$ and $\Delta_t$
 by~\eqref{piece:actual-score}
 \Comment{proposal}
 \State Draw an independent $\xi_t\sim\Lap(c_{\mathrm v}\tau\lambda)$
 \If{$\Delta_t+\xi_t\ge2\Gamma$}
 \Comment{validation}
 \State $h_{t+1}\gets\argmin_{h\in\mathcal K}
 \KL(h\Vert h_t\odot r_t^{-\theta_t})$
 \Comment{entropy projection}
 \Else
 \State $h_{t+1}\gets h_t$
 \Comment{rejection}
 \EndIf
 \State Store $h_{t+1}$
\EndFor
\State Draw $I\in\{0,\ldots,N\}$ with
 $\Pr[I=i]\propto\exp\bigl(-\eps_0V(h_i)/(2(1+a))\bigr)$
 \Comment{exponential mechanism}
\State \Return $h_I$
\end{algorithmic}
\end{algorithm}

Algorithm~\ref{piece:algorithm} assembles these steps, writing $r_t$
and $\Delta_t$ for the quantities~\eqref{piece:actual-score} of trial
$t$. It projects each accepted update back onto $\mathcal K$ in
relative entropy, which cannot increase the divergence from
$b\in\mathcal K$. The final selection score $V$ has sensitivity $1+a$,
since $\SDP$ is $1$-Lipschitz in $\ell_1$ and $\theta E-aT$ changes by
at most $(1+a)\|b-b'\|_1$.

The full mechanism releases $Y$ by~\eqref{decomp:eq-heavy}, constructs
the layers and the confidence sets~\eqref{decomp:eq-confidence}, and
calls Algorithm~\ref{piece:algorithm} in decreasing layer order, using
zero outputs in the cases specified above. It returns
the graph $H$ with edge-weight vector $w_H=Y+\sum_jh_j$.
 \section{Analysis}\label{sec:analysis}

We analyze the log-determinant oracle in Section~\ref{sec:oracle},
establish accuracy for a single layer in Section~\ref{sec:piece-accuracy},
and prove privacy of the full mechanism in Section~\ref{sec:piece-privacy}.
We then combine the layer guarantees in Section~\ref{sec:global}
to prove Theorem~\ref{thm:main}.

\subsection{The log-determinant oracle}\label{sec:oracle}

The log-determinant term makes the optimizer stable
(Lemma~\ref{oracle:lem-rdp}) at the price of the harmonic error term in
Lemma~\ref{oracle:thm-harmonic}. Throughout this subsection, we fix the
current vertex set $V$ and parameters $a,\lambda>0$, and we analyze the
optimizer $X(E,T)$ defined in~\eqref{oracle:eq-optimizer}. The diagonal
constraint ensures that $q(X)\in[0,1]^{\binom V2}$. The adjacency
objective differs from the cut objective by a degree term, which appears
explicitly in the accuracy bound below. We begin with the optimality
conditions; the spectral bounds in the next lemma are needed for the
implementation in Appendix~\ref{sec:implementation}.

\begin{lemma}[Optimality and conditioning]\label{oracle:lem-kkt}
The maximizer $X=X(E,T)$ exists and is unique. There is a vector
$\nu\ge0$ such that
\[
 X^{-1}=\diag(\nu)+\frac{aL_T+A_E}{4\lambda},\qquad
 \nu_i(X_{ii}-1)=0,\qquad (X^{-1})_{ii}\ge1.
\]
Moreover,
\[
 \Bigl(|V|+\frac{\norm{E}_1+a\norm{T}_1}{2\lambda}\Bigr)^{-1}I
 \preceq X\preceq|V|\,I.
\]
\end{lemma}
\begin{proof}
The closure of the domain is compact, and $\log\det X$ tends to
$-\infty$ at its singular boundary, so a positive definite maximizer
exists. Strict concavity gives uniqueness, and the strictly feasible
point $I/2$ gives the KKT conditions. Cauchy--Schwarz yields
$(X^{-1})_{ii}X_{ii}\ge1$, hence $(X^{-1})_{ii}\ge1$.
Multiplying the KKT equation by $X$, taking traces, and using
complementary slackness, we obtain
\[
 \tr(X^{-1})=|V|-\frac{\tr(A_EX)}{4\lambda}
 +\frac a\lambda\sum_eT_e\bigl(1/2-q_e(X)\bigr)
 \le|V|+\frac{\norm{E}_1+a\norm{T}_1}{2\lambda}.
\]
This bounds the smallest eigenvalue of $X$, and $\tr X\le|V|$ bounds
the largest.
\end{proof}

The quadratic term $a\ip{T}{q(X)^2}$ in the next lemma accounts for the
second-order loss in the analysis of the sampled update in
Lemma~\ref{piece:one-proposal}. Including this term in the oracle bound
yields the certificate~\eqref{piece:oracle-interface} for $Q_\theta$.

\begin{lemma}[Oracle accuracy]\label{oracle:thm-harmonic}
Let $\lambda>0$ and $0<a\le1$. For a signed edge vector $E$ and an edge
vector $T$ satisfying $0\le T_e\le\lambda$, let $X=X(E,T)$. Then
\[
 \SDP(E-aT)+a\ip{T}{q(X)^2}
 \le4\ip{E}{q(X)}+15\lambda^2\sum_{v\in V}\frac1{\lambda+a\deg_T(v)}
 +\frac34\norm{\deg(E)}_1.
\]
\end{lemma}

To prove the lemma, we normalize $X^{-1}$ by its diagonal $K$ and bound
a single nonnegative trace in two ways: from above by the adjacency
signal, by completing the square, and from below by the
degree-corrected SDP value, through the KKT conditions. The remaining
terms are $O(\tr(K^{-1}))$, which yields the harmonic sum.

\begin{proof}
Set $d_i=\deg_T(i)$ and
$K=\diag((X^{-1})_{ii})$. Lemma~\ref{oracle:lem-kkt} gives
$K_{ii}=\nu_i+ad_i/(4\lambda)\ge\max\{1,ad_i/(4\lambda)\}$.
We normalize by this diagonal:
\[
 Q=K^{-1/2}X^{-1}K^{-1/2}=I+B\succ0,\qquad
 R=\frac{a}{4\lambda}K^{-1/2}A_TK^{-1/2},\qquad t=\norm{R}_F^2.
\]
Both $B$ and $R$ have zero diagonal, and $\norm{R}_{\mathrm{op}}\le1$
because $|x^\top A_Tx|\le\sum_i d_i x_i^2$ and $K\succeq(a/(4\lambda))\diag(d)$.
The edge cap $T_e\le\lambda$ and the inequality
$2/(K_{ii}K_{jj})\le K_{ii}^{-2}+K_{jj}^{-2}$ give
\begin{equation}\label{oracle:eq-normalized-bounds}
 t\le\frac{a^2}{16\lambda}\sum_i\frac{d_i}{K_{ii}^2}
 \le\frac a4\tr(K^{-1})\le\frac14\tr(K^{-1}).
\end{equation}
In these coordinates, the normalized adjacency signal $F$ and the
matrix $G$ are
\[
 F\coloneqq-\frac{\tr(A_EX)}{4\lambda}=\tr\bigl((B^2+RB)Q^{-1}\bigr),
 \qquad G\coloneqq B^2Q^{-1}\succeq0,
\]
where the equality uses $\tr(B+R)=0$. We next bound
$\tr((I-R)G)$ by $2(F+t)$ through the following square completion:
\[
 4(F+t)-2\tr\bigl((I-R)G\bigr)
 =\bigl\|Q^{-1/2}[B+(Q+I)R]\bigr\|_F^2+\tr\bigl(G(I-R^2)\bigr)\ge0.
\]
To verify the identity, use $B=Q-I$, so that $B$ commutes with $Q$,
and cyclicity of the trace.

To relate the SDP value to $G$, define $M=K^{-1/2}GK^{-1/2}\succeq0$.
The identity $G=Q^{-1}-2I+Q$ gives
\[
 X\preceq M+2K^{-1},\qquad M_{ii}=X_{ii}-K_{ii}^{-1}.
\]
Let $\widehat V=\SDP(E-aT)-\frac14\sum_i\deg_E(i)$. Testing the KKT
equation against a unit-diagonal matrix gives
$\widehat V\le\lambda\sum_i\nu_i$, and complementary slackness together
with $(1-K_{ii}^{-1})^2\ge0$ implies
\[
 \nu_i\le2\nu_iM_{ii}+\nu_i/K_{ii}^2,\qquad
 \widehat V\le2\lambda\tr\bigl(\diag(\nu)M\bigr)
 +\lambda\sum_i\frac{\nu_i}{K_{ii}^2}.
\]
For the quadratic term, note that for $0\le u\le1$ and $v,w\ge0$,
$u\le v+w$ implies $u^2\le2v+w^2$ (consider $w\ge u$ and $w<u$).
Applying this with $u=q_e(X)$, $v=q_e(M)$, and $w=2q_e(K^{-1})$, we
obtain
\[
 a\ip{T}{q(X)^2}\le2a\ip{T}{q(M)}+\frac a2\sum_i\frac{d_i}{K_{ii}^2}.
\]
The two residual sums combine to at most $2\lambda\tr(K^{-1})$, since
$\nu_i+ad_i/(4\lambda)=K_{ii}$. Moreover,
\[
 \lambda\tr\bigl(\diag(\nu)M\bigr)+a\ip{T}{q(M)}
 =\lambda\tr\bigl((I-R)G\bigr).
\]
Consequently,
\begin{align*}
 \widehat V+a\ip{T}{q(X)^2}
 &\le2\lambda\tr\bigl((I-R)G\bigr)+2\lambda\tr(K^{-1})\\
 &\le4\lambda F+3\lambda\tr(K^{-1})\\
 &\le4\lambda F+15\lambda^2\sum_i\frac1{\lambda+ad_i},
\end{align*}
where the second inequality uses the completed square
and~\eqref{oracle:eq-normalized-bounds}, and the third uses
$5K_{ii}\ge1+a d_i/\lambda$.

Finally, $\lambda F=\ip{E}{q(X)}-\frac14\sum_i\deg_E(i)X_{ii}$, so
converting back to the cut objective adds
\[
 \sum_i(1/4-X_{ii})\deg_E(i)\le\frac34\norm{\deg(E)}_1,
\]
which proves the lemma.
\end{proof}

We turn to stability: Gaussian sampling from the optimizer changes little
under small changes of the input.

\begin{lemma}[Gaussian R\'enyi stability]\label{oracle:lem-rdp}
Let $X_i=X(E_i,T_i)$, $i\in\{0,1\}$, for the same $a,\lambda>0$, and put
\[
 \chi=\frac{\norm{E_1-E_0}_1/2+a\norm{T_1-T_0}_1}{\lambda}.
\]
There are absolute constants $c,C>0$ such that, for every $\alpha>1$
with $\alpha\chi\le c$,
\begin{equation}\label{oracle:eq-rdp}
 D_\alpha\bigl(N(0,X_0)\Vert N(0,X_1)\bigr)\le C\alpha\chi^2,
\end{equation}
and the same bound holds with the inputs exchanged.
\end{lemma}
\begin{proof}
Write $\Delta X=X_1-X_0$, $\Delta E=E_1-E_0$, $\Delta T=T_1-T_0$,
and $Z=X_0^{-1/2}\Delta X X_1^{-1/2}$. Adding the two first-order
optimality inequalities gives
\[
 \lambda\norm{Z}_F^2
 =\lambda\tr\bigl((X_0^{-1}-X_1^{-1})\Delta X\bigr)
 \le\Bigl|\frac14\tr(A_{\Delta E}\Delta X)
 +a\ip{\Delta T}{q(\Delta X)}\Bigr|.
\]
Factoring $\Delta X$ through $Z$ gives $|\Delta X_{uv}|\le\norm{Z}_F$
and $|q_{uv}(\Delta X)|\le\norm{Z}_F\sqrt{q_{uv}(X_0)q_{uv}(X_1)}
\le\norm{Z}_F$. Hence $\norm{Z}_F\le\chi$.

It remains to convert this bound into a R\'enyi bound. The eigenvalues
$t_i>0$ of $X_0^{-1/2}X_1X_0^{-1/2}$ satisfy
$\sum_i(t_i-1)^2/t_i=\norm{Z}_F^2\le\chi^2$. For $\chi\le1/2$, this implies
$t_i\in[1/2,2]$ and $\sum_i(t_i-1)^2\le2\chi^2$. Taking $c$ sufficiently
small ensures $|\alpha(t_i-1)|\le1/4$, so the Gaussian density integral
is finite and
\[
 D_\alpha\bigl(N(0,X_0)\Vert N(0,X_1)\bigr)
 =\frac1{2(\alpha-1)}\sum_i f_\alpha(t_i-1),\qquad
 f_\alpha(u)=\alpha\log(1+u)-\log(1+\alpha u).
\]
Since $f_\alpha(0)=0$ and
$f_\alpha'(u)=\alpha(\alpha-1)u/\bigl((1+u)(1+\alpha u)\bigr)$,
integration gives $0\le f_\alpha(u)\le C\alpha(\alpha-1)u^2$
on $|\alpha u|\le1/4$. Substitution proves~\eqref{oracle:eq-rdp}, and
exchanging the inputs gives the reverse direction.
\end{proof}

In a layer, the sign $\theta$ and the public quantities $h$ and $P$ are
fixed, and the private input enters only through $E=\theta(h-b)$ and
$T=h+b+P$. The perturbation parameter of Lemma~\ref{oracle:lem-rdp} is
therefore $\chi=(1/2+a)\norm{b-b'}_1/\lambda$.

\subsection{Accuracy of a layer}\label{sec:piece-accuracy}

We analyze a single layer in the notation of Section~\ref{sec:piece}.
The next lemma shows that the sampled score of a Gaussian proposal is
at least $\tau$ times its covariance score with constant probability,
and that it equals the exact entropy decrease before projection.

\begin{lemma}[Sampled update]\label{piece:one-proposal}
Fix a candidate $h$ positive on the relative support of $\mathcal K$
and a sign $\theta$. The sampled score in~\eqref{piece:actual-score}
satisfies $\Pr[\Delta_\theta\ge\tau Q_\theta(h)]\ge1/120$.
For arbitrary edge factors $r_e\in[1,2]$, define
$\Delta_\theta$ by the same formula and let
$h^+=\argmin_{f\in\mathcal K}\KL(f\Vert h\odot r^{-\theta})$.
Then every $f\in\mathcal K$ satisfies
\begin{equation}\label{piece:progress}
 \KL(f\Vert h^+)-\KL(f\Vert h)
 \le-\Delta_\theta+(\log2)\|f-b\|_1.
\end{equation}
In particular, if $b\in\mathcal K$, the entropy $\KL(b\Vert\cdot)$
decreases by at least $\Delta_\theta$.
\end{lemma}
\begin{proof}
For $x\ge0$ and $r=(1+2x)/(1+x)$, we have
\[
 x-\frac32x^2\le\log r=\int_x^{2x}\frac{dt}{1+t}\le x,
 \qquad 1-r^{-1}\ge x-2x^2,\qquad r-1\le x.
\]
Applying these inequalities with $x_e=\tau g_e$ gives
\[
 \Delta_+\ge\tau\ip{h-b}{g}-2\tau^2\ip{h}{g^2},\qquad
 \Delta_-\ge\tau\ip{b-h}{g}-\frac32\tau^2\ip{b}{g^2}.
\]
Put $\mu=q(X_\theta)$, $F=\theta\ip{h-b}{g}$,
$Z=\ip{T}{g^2}$, and $B=\ip{T}{\mu^2}$. Since $h,b\le T$,
\[
 \Delta_\theta\ge\tau F-2\tau^2Z,\qquad
 \E F=Q_\theta+\frac a4B,\qquad \E Z=3B.
\]
The centered quadratic form $F_0=F-\E F$ has the law
$\sum_i c_i(G_i^2-1)$ for independent standard normal $G_i$. Hence
\[
 v^2\coloneqq\E F_0^2=2\sum_i c_i^2,\qquad
 \E F_0^4=12\Bigl(\sum_i c_i^2\Bigr)^2+48\sum_i c_i^4\le15v^4.
\]
If $v>0$, interpolation gives $\E|F_0|\ge v/\sqrt{15}$. Since
$\E(F_0)_+=\E|F_0|/2$, Cauchy--Schwarz implies
$\Pr[F_0>0]\ge(\E(F_0)_+)^2/\E(F_0)_+^2\ge1/60$.
If $v=0$, then $F=\E F$ almost surely. If $B>0$, Markov's inequality
gives $\Pr[Z>360B]\le1/120$. Thus $F\ge\E F$ and $Z\le360B$
hold together with probability at least $1/120$; since $720\tau=a/4$,
they imply $\Delta_\theta\ge\tau Q_\theta$.
If $B=0$, every edge charged by $T$ has $g_e=0$ almost surely, and
both scores vanish.

For the entropy bound, put $u=h\odot r^{-\theta}$. The initializer
and the entropy projections keep the iterates positive on the relative
support. The unnormalized entropy identity gives
\[
 \KL(f\Vert u)-\KL(f\Vert h)
 =\theta\ip{f}{\log r}+\ip{h}{r^{-\theta}-1}
 =-\Delta_\theta+\theta\ip{f-b}{\log r}.
\]
Bregman projection onto $\mathcal K$ cannot increase the divergence from
any $f\in\mathcal K$. Since $0\le\log r_e\le\log2$, this proves
\eqref{piece:progress}, including the case $f=b$.
\end{proof}

When validation is accurate, the entropy bound~\eqref{piece:progress}
limits the number of accepted updates, while at a candidate with a large
violation each trial is accepted with constant probability. Hence, with
high probability, an accurate candidate appears within the horizon, and
the exponential mechanism selects a nearly accurate one.

\begin{lemma}[Accuracy of a layer]\label{piece:release}
In the layer setup of Section~\ref{sec:piece}, suppose that
$b\in\mathcal K$ and that~\eqref{piece:oracle-interface} holds for every
candidate and sign. With probability at least $1-\zeta$,
Algorithm~\ref{piece:algorithm} returns a graph $\widehat h$ satisfying
\begin{equation}\label{piece:release-bound}
 V(\widehat h)\le A+12\Gamma/\tau
 +\frac{2(1+a)}{\eps_0}\log\frac{3(N+1)}\zeta.
\end{equation}
Consequently, simultaneously for every cut $S$,
\begin{equation}\label{piece:cut-certificate}
 |\widehat h(S)-b(S)|
 \le a\bigl[\widehat h(S)+b(S)+P(S)\bigr]
 +O\Bigl(A+(\lambda+\eps_0^{-1})\log\frac{N+1}\zeta\Bigr).
\end{equation}
All implicit constants are absolute.
\end{lemma}
\begin{proof}
Except with probability $\zeta/6$, all validation noises have magnitude
at most $c_{\mathrm{v}}\tau\lambda\ell_N$. On this event, acceptance gives
$\Delta_t\ge\Gamma\ge\tau A$. By~\eqref{piece:progress}, each accepted
step decreases $\Phi(h)=\KL(b\Vert h)$ by at least $\tau A$, so
\eqref{piece:initial-potential} and $\Phi\ge0$ bound the number of
acceptances by $J_0=M\log(2n^2)/(\tau A)$.

Call $h$ good if $V(h)\le A+12\Gamma/\tau$.
By~\eqref{piece:oracle-interface}, at a bad candidate one sign has
$Q_\theta>3\Gamma/\tau$. By Lemma~\ref{piece:one-proposal}, the event
that the trial chooses this sign, obtains $\Delta_\theta\ge3\Gamma$, and
draws $\xi_t\ge-\Gamma$ has conditional probability at least
$p_0=(1/2)(1/120)(1/2)=1/480$. This statement conditions only on the
past. Let $\widetilde I_t$ be the acceptance indicator of trial $t$,
padded with ones after the first good candidate; the conditional means
of these padded indicators are at least $p_0$. A sufficiently large
$C_N$ in~\eqref{piece:horizon} ensures $J_0\le p_0N/2$ and
\[
 \Pr\Bigl[\sum_{t=0}^{N-1}\widetilde I_t\le p_0N/2\Bigr]
 \le e^{-p_0N/8}\le\zeta/6.
\]
If no stored candidate is good and all validation noises are small,
the sum counts actual acceptances and is therefore at most
$J_0\le p_0N/2$. Thus a good candidate is stored except with probability
$\zeta/3$. Conditional on the list, the exponential mechanism adds the
last term of~\eqref{piece:release-bound}, with failure probability at
most $\zeta/3$. Every cut is feasible for $\SDP$; substituting the value
of $\Gamma$ from~\eqref{piece:threshold} therefore
proves~\eqref{piece:cut-certificate}.
\end{proof}

\subsection{Privacy of the mechanism}\label{sec:piece-privacy}

The private input enters a trial through the covariance $X_\theta$,
which Lemma~\ref{oracle:lem-rdp} handles, and through the part
$-\theta\ip{b}{\log r}$ of the sampled score. The bound
$\log r_e\le\tau g_e$ matches the sensitivity of the latter to the
noise scale $c_{\mathrm{v}}\tau\lambda$ after averaging over the Gaussian
sample, so we analyze the sample and the validation jointly.

\begin{lemma}[Privacy of a trial]\label{privacy:trial}
Conditional on any common public history, let $s=\|b-b'\|_1\le1$.
For $\alpha>1$ and $\alpha/\lambda\le c$, a complete trial satisfies
\[
 D_\alpha(\mathrm{trial}_b\Vert\mathrm{trial}_{b'})\le C\alpha s^2/\lambda^2
\]
in either direction. This holds for the transcript consisting of the
sign, the Gaussian vector, the full noisy validation, and the updated
candidate, including rejected proposals. The covariance is not released,
and neither input needs to belong to $\mathcal K$.
\end{lemma}

By Cauchy--Schwarz, the likelihood moment of the joint release splits
into a Gaussian factor, which Lemma~\ref{oracle:lem-rdp} controls, and a
Laplace factor. We control the location shift of the Laplace factor
through exponential moments of a weighted average $g_\pi$ of the
coordinates of $g$.

\begin{proof}
Fix the sign; the case $s=0$ is immediate. Let $\mathsf P,\mathsf Q$
be the Gaussian laws under $b,b'$. Lemma~\ref{oracle:lem-rdp} applies
with perturbation $(1/2+a)s/\lambda$. Put $\pi_e=|b_e-b'_e|/s$
and $g_\pi=\sum_e\pi_eg_e$, an average of the coordinates of $g$ with weights $\pi_e\ge0$ summing to one. The standardized shift of the conditional
Laplace location satisfies
\[
 x(z)=\frac{|\Delta_\theta(z;b)-\Delta_\theta(z;b')|}
 {c_{\mathrm{v}}\tau\lambda}
 \le\frac{\sum_e|b_e-b'_e|\log r_e}{c_{\mathrm{v}}\tau\lambda}
 \le\frac{sg_\pi}{\lambda}.
\]

We first bound the Laplace factor at order $p=2\alpha-1$, the order
arising from Cauchy--Schwarz below. For equal-scale Laplace laws at
standardized distance $x\ge0$, direct integration gives the likelihood
moment of order $p$,
\[
 M_p(x)=\frac{p}{2p-1}e^{(p-1)x}+\frac{p-1}{2p-1}e^{-px}
 \le1+\frac12p(p-1)x^2e^{(p-1)x},
\]
where the inequality follows from $M_p(0)=1$, $M_p'(0)=0$, and
$M_p''(x)\le p(p-1)e^{(p-1)x}$. Under either Gaussian law, $g_e$ has
distribution $\mu_eZ^2$, where $Z\sim N(0,1)$ and $0\le\mu_e\le1$.
Convexity of $u^2e^{ku}$ and Jensen's inequality give
\[
 \E g_\pi^2e^{kg_\pi}\le\sum_e\pi_e\E g_e^2e^{kg_e}
 \le\E Z^4e^{kZ^2}=3(1-2k)^{-5/2}\qquad(0\le k<1/2).
\]
Consequently, if $(p-1)/\lambda\le1/4$, then
\begin{equation}\label{privacy:laplace-moment}
 \E_{\mathsf P}M_p\bigl(x(z)\bigr)\le\exp\{Cp(p-1)s^2/\lambda^2\}.
\end{equation}

We now combine the two factors. Let $\Lambda_{\mathrm G}=d\mathsf P/d\mathsf Q(z)$, and
let $\Lambda_{\mathrm V}$ be the likelihood ratio of the conditional Laplace laws. With
expectations under the joint law of the first input, Cauchy--Schwarz
gives
\[
 \E(\Lambda_{\mathrm G}\Lambda_{\mathrm V})^{\alpha-1}
 \le\bigl(\E\Lambda_{\mathrm G}^{p-1}\bigr)^{1/2}\bigl(\E\Lambda_{\mathrm V}^{p-1}\bigr)^{1/2}
 =\exp\{(\alpha-1)D_p(\mathsf P\Vert\mathsf Q)\}
 \bigl(\E_{\mathsf P}M_p(x(z))\bigr)^{1/2}.
\]
The left side is the order-$\alpha$ likelihood moment of the joint
Gaussian and validation release. Applying the Gaussian stability bound
and~\eqref{privacy:laplace-moment}, taking logarithms, and dividing by
$\alpha-1$ gives $D_\alpha\le C\alpha s^2/\lambda^2$. Since
$p\le2\alpha$, a sufficiently small $c$ ensures both moment conditions.
Exchanging the inputs gives the reverse bound. Finally, the sign is
independent of the data, and the updated candidate is postprocessing.
\end{proof}

To compose the trials, we need four properties of the
schedule~\eqref{global:eq-parameters}, which are also used in
Section~\ref{sec:global}. Let $N_{\mathrm{tot}}$ denote the total number
of trials over all layers.

\begin{lemma}[Schedule bounds]\label{global:lem-schedule}
On every execution, for every layer $j$,
\begin{equation}\label{global:eq-schedule}
 \lambda_j\ge c\sigma,\quad
 \frac{N_j}{\lambda_j^2}=O(\eps/\sigma),\quad
 N_{\mathrm{tot}}=O\bigl(\eps\sigma[J+1+n/\gamma]\bigr),\quad
 \log\frac{(N_j+1)(J+1)}{\beta}=O(\eps\sigma),
\end{equation}
where the implicit constants are independent of $C_\sigma$.
\end{lemma}
\begin{proof}
The first bound follows from~\eqref{global:eq-parameters}, since
$\lambda_j\ge\lambda_0=\sqrt{C_D}\,\sigma/\gamma$ and $\gamma\le1/4$.
For the second, the bounds $M_j\le2n_jD_j$ and $\omega_j\ge n_j$ give
\[
 \lambda_j^2A_j\ge C_A\sigma^2D_j\omega_j/\gamma
 \ge c\sigma^2M_j/\gamma.
\]
Together with $\lambda_j\ge c\sigma$ and~\eqref{piece:horizon}, this
gives $N_j/\lambda_j^2=O(\eps/\sigma)$ and
$N_j=O(\eps(\sigma+D_j))$. The geometric sum
$\sum_{j=1}^JD_j\le CnU$ includes an empty sum when $J=0$.
For layer zero, $A_0\ge cn\sigma/\gamma$ and $M_0\le Cn^2U$ give
$N_0=O(\eps\sigma[1+n/\gamma])$ directly. Summing proves the bound on
$N_{\mathrm{tot}}$, whose logarithm is $O(\eps\sigma)$, and the last
bound follows.
\end{proof}

The first bound keeps the R\'enyi order chosen below in the range of
Lemma~\ref{privacy:trial}, and the second makes the $N_j$ trials of
layer~$j$ cost $O(\alpha\eps s_j^2/\sigma)$. The third bound enters the
running time, and the last makes the logarithmic factor of
Lemma~\ref{piece:release}, at $\zeta=\beta_0/(J+1)$, of order
$\eps\sigma$.

\begin{proposition}[Privacy of the mechanism]\label{global:privacy}
The full mechanism is $(\eps,\delta)$-differentially private.
\end{proposition}
\begin{proof}
Fix a common preprocessing output. The distances $s_j$ and the
schedules $N_j,\lambda_j$ are then fixed, whereas the priors may depend
on earlier releases. Lemma~\ref{privacy:trial} and adaptive R\'enyi
composition bound the divergence of the trials in layer $j$ by
$C\alpha N_js_j^2/\lambda_j^2$, including rejected trials.
Conditional on the candidate list, $V$ has sensitivity $(1+a)s_j$,
so the final selection has pure privacy cost $\eps_0s_j$. Since
$\sum_js_j\le1$ by~\eqref{decomp:eq-partition},
\[
 D_\alpha\le O\Bigl(\frac{\alpha\eps}{\sigma}\Bigr)\sum_js_j^2
 +\eps_0\sum_js_j
 \le O\Bigl(\frac{\alpha\eps}{\sigma}\Bigr)+\eps_0.
\]
This holds in both directions, at every preprocessing output and
subsequent history. We choose $\alpha=1+4\log(2/\delta)/\eps$.
Equation~\eqref{global:eq-schedule} ensures $\alpha/\lambda_j\le c$.
R\'enyi conversion with failure parameter $\delta/2$ costs $\eps/4$,
and taking $C_\sigma$ sufficiently large bounds the trial term by
another $\eps/4$. The three pure-DP preprocessing releases cost
$3\eps_0$, so the ideal mechanism is $(9\eps/10,\delta/2)$-DP.
Appendix~\ref{sec:implementation} uses the remaining allowance for the
numerical implementation.
\end{proof}

\subsection{Proof of the upper bound}\label{sec:global}

We now combine the layers and prove Theorem~\ref{thm:main}. The proof
first establishes a bound that holds for all $n\ge2$, $0<\gamma\le1/4$,
$0<\eps\le1$, and $0<\delta,\beta<1/10$, with no lower bounds on these
parameters: the mechanism of Section~\ref{sec:algorithm} is
$(\eps,\delta)$-differentially private and, with probability at least
$1-\beta$, its output $H$ on input $G$ satisfies simultaneously for every
$S\subseteq V=[n]$
\begin{equation}\label{global:full-bound}
 \begin{aligned}
 |w_G(S,V\setminus S)-w_H(S,V\setminus S)|
 &\le\gamma w_G(S,V\setminus S)\\
 &\quad+O\Bigl(\frac{n}{\eps\gamma}\log\frac{n}{\gamma\eps\delta\beta}
 +\frac{1+\sqrt{n\gamma}}{\eps\gamma}
 \log^2\frac{n}{\gamma\eps\delta\beta}\Bigr),
 \end{aligned}
\end{equation}
with an absolute constant in $O(\cdot)$. The lower bounds
$\gamma,\eps,\delta,\beta\ge n^{-c}$ of Theorem~\ref{thm:main} are used
only at the end, to simplify~\eqref{global:full-bound}
to~\eqref{intro:full-bound}.

The main step is to verify the
certificate~\eqref{piece:oracle-interface} in each layer. Here the prior
supplies the degree that keeps the harmonic sum of
Lemma~\ref{oracle:thm-harmonic} small (Section~\ref{sec:overview}).

\begin{lemma}[The certificate in each layer]\label{global:lem-certificate}
Let $C_A$ be sufficiently large. On the accuracy event of
Lemma~\ref{decomp:interface}, the
certificate~\eqref{piece:oracle-interface} holds in every layer~$j$,
that is, with $b=b_j$, $P=P_j$, $\lambda=\lambda_j$, and $A=A_j$, for
every $h\in\mathcal K_j$ and both signs $\theta$.
\end{lemma}
\begin{proof}
On this event, every released layer lies in its confidence set and has
edge weights at most $U$. Since $P_j\le3U$ coordinatewise as well, we
have $T=h+b_j+P_j\le5U\le\lambda_j$ and
$\|\deg(h-b_j)\|_1\le2R_j$. The degree floors of the confidence sets,
applied to $h$ when $k=j$ and to the summand $(3/4)^{k-j}h_k$ of $P_j$
when $k>j$, show that for $v\in L_k$ with $k\ge j\ge1$,
\[
 \deg_T(v)\ge
 \begin{cases}
 D_j/4, & k=j,\\
 (3/4)^{k-j}D_k/4, & k>j.
 \end{cases}
\]
Since $D_k=2^{k-j}D_j$, the second bound equals $(3/2)^{k-j}D_j/4$;
this is the source of the weights $(2/3)^{k-j}$ in $\omega_j$.
Consequently,
\[
 \sum_{v\in V_j}\frac1{\lambda_j+a\deg_T(v)}\le\frac{C\omega_j}{aD_j}
 \quad(j\ge1),\qquad
 \sum_{v\in V_0}\frac1{\lambda_0+a\deg_T(v)}\le\frac n{\lambda_0}.
\]
Since $\lambda_j^2=\sigma D_j$ and
$\lambda_0=\Theta(\sigma/\gamma)$, the harmonic term in
Lemma~\ref{oracle:thm-harmonic} is $O(\sigma\omega_j/\gamma)$, and its
degree term is $O(R_j)$. As
$4Q_\theta(h)=4\theta\ip{E}{q(X_\theta)}-a\ip{T}{q(X_\theta)^2}$,
Lemma~\ref{oracle:thm-harmonic}, applied to $\theta E$ and $T$,
therefore gives~\eqref{piece:oracle-interface} with the allowance
in~\eqref{global:eq-allowance} for sufficiently large $C_A$.
\end{proof}

\begin{proof}[Proof of Theorem~\ref{thm:main}]
We first establish~\eqref{global:full-bound}. Privacy follows from
Proposition~\ref{global:privacy}. For accuracy, we work on the accuracy
event of Lemma~\ref{decomp:interface}. On this event, $b_j\in\mathcal K_j$
for every $j$, and Lemma~\ref{global:lem-certificate} provides the
certificate in every layer. Lemma~\ref{piece:release} therefore applies
in every layer with $\zeta=\beta_0/(J+1)$, and its logarithmic factor
is $O(\eps\sigma)$ by~\eqref{global:eq-schedule}. Hence, simultaneously
for all layers and cuts,
\[
 |h_j(S)-b_j(S)|\le a\bigl[h_j(S)+b_j(S)+P_j(S)\bigr]+B_j,
 \qquad B_j=O(A_j+\eps\sigma\lambda_j+\sigma).
\]
The zero outputs are exact on this event, and the layer failure
probabilities sum to at most $(J+1)\zeta=\beta_0$.

It remains to sum over the layers. The sums $\sum_j\omega_j\le4n$ and
$\sum_jR_j=O(n\sigma)$ give $\sum_jA_j=O(n\sigma/\gamma)$, and the
geometrically growing regularization parameters satisfy
$\sum_j\lambda_j\le C\sigma\bigl(\gamma^{-1}+\sqrt{n/\gamma}\bigr)$. It
follows that
\begin{equation}\label{global:eq-summed-error}
 \sum_jB_j=O\Bigl(\frac{\sigma}{\gamma}
 \bigl[n+\eps\sigma(1+\sqrt{n\gamma})\bigr]\Bigr).
\end{equation}
The relative slacks add up as well: with $h_{\mathrm{c}}=\sum_jh_j$ and
$\sum_jP_j\le3h_{\mathrm{c}}$, we obtain
\[
 |h_{\mathrm{c}}(S)-b(S)|
 \le a\bigl[4h_{\mathrm{c}}(S)+b(S)\bigr]+\sum_jB_j.
\]
Rearranging gives a relative excess of at most $5a/(1-4a)\le\gamma/2$,
a relative deficit no larger than this, and an absolute constant
multiplying $\sum_jB_j$.

Finally, the heavy-edge release contributes at most $\gamma w(S)/2$,
with failure probability at most $\beta_0$, and $b(S)\le w(S)$.
Substituting $\sigma$ from~\eqref{global:eq-parameters}
into~\eqref{global:eq-summed-error} proves~\eqref{global:full-bound}.
The preprocessing, layer, and numerical failure probabilities fit within
$\beta$.

Under the hypotheses of Theorem~\ref{thm:main},
$L=\log(n/(\gamma\eps\delta\beta))\le(4c+1)\log n$.
Since $1+\sqrt{n\gamma}\le2\sqrt n$ and $\log n\le\sqrt n$, we have
$(1+\sqrt{n\gamma})L^2=O(n\log n)$, with a constant depending only on
$c$. Thus~\eqref{global:full-bound} implies~\eqref{intro:full-bound}.

For the running time, there are
$O((n/\gamma)\log(n/(\gamma\eps\delta\beta)))$ private trials and
$O(N_{\mathrm{tot}}+J+1)$ final SDP evaluations. Each trial uses a
log-determinant optimization, a Gaussian sample, and a feasible entropy
update. Appendix~\ref{sec:implementation} implements these operations
in time polynomial in the input encoding length, $n$, $1/\gamma$,
$1/\eps$, $\log(1/\delta)$, and $\log(1/\beta)$, which is polynomial in
$n$ and the input encoding length under the hypotheses of
Theorem~\ref{thm:main}.
\end{proof}
 \section{Lower bound}\label{sec:lower}

We prove Theorem~\ref{thm:intro-lower} by establishing a more general
bound. Let $n$ be sufficiently large, $0<\eps\le1$,
$0\le\delta\le\eps/100$, $0\le\beta\le1/4$, and $0<\gamma\le1/4$, and
consider an $(\eps,\delta)$-differentially private mechanism that, on
every nonnegative weighted input graph $G$ on $V=[n]$, outputs a nonnegative weighted
graph $H$ satisfying~\eqref{intro:guarantee} simultaneously for every
cut with probability at least $1-\beta$. We show that
\begin{equation}\label{lower:general}
 B=\Omega\Bigl(\frac{n}{\eps}\min\{\sqrt n,\gamma^{-1}\}
   \min\bigl\{\log(2+n\gamma^2),\log(\eps/\delta)\bigr\}\Bigr),
\end{equation}
with an absolute implied constant and $\log(\eps/\delta)=+\infty$ when
$\delta=0$. Compared with Theorem~\ref{thm:intro-lower}, this bound
holds for every relative error $\gamma\le1/4$, and its logarithmic
factor is capped at $\log(\eps/\delta)$, reflecting the limited reach of
group privacy when $\delta>0$.

The proof follows the reconstruction outline at the end of
Section~\ref{sec:overview}. We run the mechanism on a random sparse
bipartite graph whose edges have a common weight $W$ and decode from its
output a graph that is consistent with it on every cut. Privacy prevents
the decoder from recovering most edges, and random signs turn the missed
edges into a cut on which the input and the decoded graph differ by more
than the relative term can absorb.

Throughout, we fix parameters and a mechanism as above. Put
$\bar n=\lfloor n/2\rfloor$ and
\[
 d=\Bigl\lfloor\min\Bigl\{\frac{\bar n}{256},\frac{1}{16\gamma^2}\Bigr\}\Bigr\rfloor,
 \qquad
 \ell=\min\Bigl\{\log\frac{\bar n}{d},\log\frac{\eps}{\delta}\Bigr\},
 \qquad
 W=\Bigl\lfloor\frac{\ell}{4\eps}\Bigr\rfloor,
\]
with the second logarithm infinite when $\delta=0$. For sufficiently
large $n$, these choices give $d\ge1$, $\gamma\sqrt d\le1/4$, and
$\ell/(8\eps)\le W\le\ell/(4\eps)$.

Let $\mathcal F_d$ consist of the simple bipartite graphs on two fixed
$\bar n$-vertex sets in which every left vertex has degree exactly $d$.
Right degrees are unrestricted, and any vertex outside the two sets is
isolated. For $G\in\mathcal F_d$, let $w_G$ be its $0$--$1$ edge vector,
and write $WG$ for the graph with edge weights $Ww_G$. We draw $G$
uniformly from $\mathcal F_d$ and run the mechanism on $WG$, whose
$m=\bar nd$ edges each have weight $W$.
From the output $H$, we decode the first $F\in\mathcal F_d$, in a fixed
order, satisfying
\begin{equation}\label{lower:compatible}
 |Ww_F(S,V\setminus S)-w_H(S,V\setminus S)|
 \le\gamma Ww_F(S,V\setminus S)+B
 \quad\text{for all }S\subseteq V,
\end{equation}
or a fixed member of $\mathcal F_d$ if no such graph exists. This
decoder need not be efficient, and it always returns a graph with $m$
edges. We write $|G\cap F|=|E(G)\cap E(F)|$ for the number of correctly
recovered edges, and we let $\mathcal U$ be the accuracy event for the
input $WG$, so that $\Pr[\mathcal U]\ge1-\beta$. On $\mathcal U$,
$G$ is compatible with $H$, so the decoded graph $F$ is compatible as well.

The first claim bounds the expected fraction of recovered edges,
without conditioning on the output being accurate.

\begin{claim}\label{lower:claim-privacy}
Over the choice of $G$ and the randomness of the mechanism,
$\E|G\cap F|<m/16$.
\end{claim}

\begin{proof}
The ratio $\E|G\cap F|/m$ is the probability that a uniform edge of $G$
belongs to $F$. Delete a uniform edge $(i,j)$ of $G$. Conditional on
its left endpoint $i$ and the remaining graph $G^-$, the missing
neighbor is uniform among $\bar n-d+1$ positions by symmetry.
On input $WG^-$, the decoded neighbor set of $i$
has size $d$ and is independent of the missing position, so it
contains that position with probability at most $d/(\bar n-d+1)$. The
inputs $WG^-$ and $WG$ differ by $W$ unit weight changes, and group
privacy (Section~\ref{sec:preliminaries}) therefore gives
\[
 \frac{\E|G\cap F|}{m}
 \le e^{\eps W}\frac{d}{\bar n-d+1}+\delta\,\frac{e^{\eps W}-1}{e^\eps-1}
 \le\frac{(d/\bar n)^{3/4}}{1-d/\bar n}+(\delta/\eps)^{3/4}
 <\frac{1}{16}.
\]
The second inequality uses
$e^{\eps W}\le e^{\ell/4}=\min\{(\bar n/d)^{1/4},(\eps/\delta)^{1/4}\}$
and $e^\eps-1\ge\eps$, and the last uses $d/\bar n\le1/256$ and
$\delta/\eps\le1/100$. The additive privacy term is zero when
$\delta=0$.
\end{proof}

The next claim turns missed edges into a cut discrepancy after
subtracting the relative error allowance.

\begin{claim}\label{lower:claim-discrepancy}
For every $G,F\in\mathcal F_d$, there is a cut $S\subseteq[n]$ such that
\[
 w_G(S,V\setminus S)-w_F(S,V\setminus S)
 -\gamma\bigl(w_G(S,V\setminus S)+w_F(S,V\setminus S)\bigr)
 \ge\frac{2(m-|G\cap F|)}{5\sqrt d}-\gamma m.
\]
\end{claim}

\begin{proof}
Let $A=G-F$ and $Q=G+F$ be the difference and the sum of the bipartite
adjacency matrices of $G$ and $F$, with rows indexed by the left
vertices. Draw independent fair signs $y_j$ on the right vertices, set
$x_i=-\operatorname{sign}((Ay)_i)$ on the left vertices with independent
fair ties, and let $S$ be the set of vertices with positive sign. An
edge $ij$ is cut exactly when $x_iy_j=-1$, so
$w_G(S,V\setminus S)=\frac12(m-x^\top Gy)$,
and likewise for $F$. If row $i$ of $A$ has $k_i$ nonzeros, then
$k_i\le2d$ and $\sum_ik_i=2(m-|G\cap F|)$. The variable $(Ay)_i$ has
second moment $k_i$ and fourth moment at most $3k_i^2$, so H\"older
interpolation gives
\[
 \E|(Ay)_i|\ge\sqrt{k_i/3}\ge\frac{2k_i}{5\sqrt d},
\]
including for zero rows.

It remains to show that $\E[x_i(Qy)_i]=0$ for every left vertex $i$.
The positive and negative supports of row $i$ of $A$ have equal size,
since $G$ and $F$ both have left degree $d$. Exchanging the signs $y_j$
within pairs of positions from these two supports negates $(Ay)_i$ and
preserves $(Qy)_i$. This exchange preserves the law of the signs, and
together with the fair ties it gives $\E[x_i(Qy)_i]=0$. Since
$x_i(Ay)_i=-|(Ay)_i|$, the cut identities therefore yield
\begin{align*}
 \E_S[w_G(S,V\setminus S)-w_F(S,V\setminus S)]
 &=\frac12\sum_i\E|(Ay)_i|\ge\frac{2(m-|G\cap F|)}{5\sqrt d},\\
 \E_S[w_G(S,V\setminus S)+w_F(S,V\setminus S)]
 &=m-\frac12\sum_i\E[x_i(Qy)_i]=m.
\end{align*}
Subtracting $\gamma$ times the second expectation from the first shows
that some cut satisfies the claimed inequality.
\end{proof}

\begin{proof}[Proof of~\eqref{lower:general}]
On $\mathcal U$, both $G$ and $F$ satisfy~\eqref{lower:compatible}, so
the triangle inequality gives
\[
 W\bigl[w_G(S,V\setminus S)-w_F(S,V\setminus S)
 -\gamma\bigl(w_G(S,V\setminus S)+w_F(S,V\setminus S)\bigr)\bigr]\le2B
 \quad\text{for every cut }S.
\]
Applying Claim~\ref{lower:claim-discrepancy}, multiplying by
$\one_{\mathcal U}$, and taking expectations over $G$ and the mechanism
gives
\[
 2B\ge\frac{2W}{5\sqrt d}\bigl(m(1-\beta)-\E|G\cap F|\bigr)-\gamma mW
 \ge\frac{mW}{40\sqrt d},
\]
where we use $\Pr[\mathcal U]\ge1-\beta$ and
$\E[\one_{\mathcal U}|G\cap F|]\le\E|G\cap F|$, followed by
Claim~\ref{lower:claim-privacy}, $\beta\le1/4$, and
$\gamma\sqrt d\le1/4$. Since $m=\bar nd$, we obtain
$B\ge\bar nW\sqrt d/80$.

It remains to evaluate this bound. Since
$t/2\le\lfloor t\rfloor\le t$ for $t\ge1$, the choice of $d$ gives
\[
 \sqrt d=\Theta\bigl(\min\{\sqrt n,\gamma^{-1}\}\bigr),
 \qquad
 \log\frac{\bar n}{d}=\Theta\bigl(\log(2+n\gamma^2)\bigr).
\]
Together with $W\ge\ell/(8\eps)$, these estimates turn
$B\ge\bar nW\sqrt d/80$ into~\eqref{lower:general}.
\end{proof}

\begin{proof}[Proof of Theorem~\ref{thm:intro-lower}]
For sufficiently large $n$, the theorem's hypotheses imply
$\delta\le\eps/100$, so~\eqref{lower:general} applies. Since
$\gamma\ge n^{-1/2+\kappa}$ and $\delta\le\eps/n^2$, the first minimum
in~\eqref{lower:general} equals $\gamma^{-1}$, while both logarithms
in the second are at least $2\kappa\log n$. This proves
\eqref{intro:lower}, with a constant depending only on $\kappa$.
\end{proof}

Theorems~\ref{thm:main} and~\ref{thm:intro-lower} give the optimal
additive error $\Theta(n\log n/(\eps\gamma))$ under their common
parameter assumptions, with constants depending only on $c$ and
$\kappa$. The same conclusion extends to $\beta\le1/4$ by running the
upper-bound algorithm with failure parameter $\min\{\beta,1/20\}$.
More generally, for fixed $\gamma,\eps$ and
$\delta=n^{-c_\delta}$, $\beta=n^{-c_\beta}$ with fixed
$c_\delta,c_\beta>0$, \eqref{lower:general} and
Theorem~\ref{thm:main} give the optimal error $\Theta(n\log n)$.

\paragraph{AI Disclosure.}
The main results and proofs presented here were obtained in August 2026
through interactions with GPT-5.6 Sol.
After becoming aware of the work of Eli\'{a}\v{s} et al.~\cite{EliasEtAl26},
we used GPT-6 Astra to improve the exposition and to assist with the
\href{https://github.com/liuchliuch/dp-cut-release-lean}{\pinkfoldmark\nobreak\hspace{0.2em}Lean formalization}.
The authors take full responsibility for the mathematical claims and the final manuscript.
\setlength{\abovedisplayskip}{11pt plus 3pt minus 6pt}
\setlength{\belowdisplayskip}{11pt plus 3pt minus 6pt}
\bibliographystyle{alpha}
\newcommand{\etalchar}[1]{$^{#1}$}

\clearpage
\appendix
\section{Implementation}\label{sec:implementation}

The analysis of Section~\ref{sec:analysis} concerns an ideal mechanism
that computes exact entropy projections and log-determinant optimizers
and draws continuous Gaussian and Laplace samples. We sketch a
polynomial-time finite-precision implementation that
satisfies~\eqref{global:full-bound}, and hence Theorem~\ref{thm:main}.
Two points need care: approximate entropy projections must still make
progress, and approximate sampling must not invalidate the privacy
analysis. We handle the first directly (Section~\ref{sec:impl-updates})
and the second by coupling the implementation to a quantized reference
mechanism in total variation (Section~\ref{sec:impl-coupling}).

Throughout, candidates and bounded inputs have mass at most
$\overline M=Cn^2U$ on every execution, and priors have mass at most
$3\overline M$. We use rational public parameters satisfying the
calibration inequalities; in particular, we approximate
$\lambda_j=\sqrt{\sigma D_j}$ within a constant factor while retaining
$\lambda_j\ge5U$, and absorb this factor in $C_A$.
Equation~\eqref{global:eq-schedule} bounds the number of trials by a
polynomial. We fix a public polynomial bound $Q_*\ge1+Q_{\mathrm c}$
on the number of sampling and score calls on every execution.

\subsection{Approximate entropy updates}\label{sec:impl-updates}

To approximate the entropy gradient uniformly, we keep candidates
bounded away from zero on the relative support by restricting each
nonempty domain to
$\mathcal K_\psi=\{h\in\mathcal K:h_e\ge\psi(h_0)_e\text{ for all }e\}$
with $\psi=c_\psi\gamma/n^2$. The private block need not lie in
$\mathcal K_\psi$, so the analysis uses the comparator
$b^\psi=(1-\psi)b+\psi h_0$, while the oracle and the scores still use
$b$. On the accuracy event, $b^\psi\in\mathcal K_\psi$ and
$\|b^\psi-b\|_1\le2\psi M$; the bounds $A_j\ge c\sigma/\gamma$ and
$M_j\le Cn^2\sigma/\gamma$ give $2\psi M_j\le c\tau A_j$ for
sufficiently small $c_\psi$, and convexity preserves the initial entropy
bound~\eqref{piece:initial-potential} for $b^\psi$.

Set $\ell=\psi\min_{e:(h_0)_e>0}(h_0)_e>0$, so the coordinates of every
$h\in\mathcal K_\psi$ on the relative support lie in $[\ell,U]$.
Let $\mu=\min\{1/8,c\tau A/M\}$. On a geometric mesh of $[\ell,U]$
with ratio $1+\mu/4$, approximate $\log t$ by rational values within
$\mu/4$ and take their running maxima. Integrating the resulting
nondecreasing step function gives a rational convex piecewise-linear
function $\varphi$ with $O((1+M/(\tau A))[1+\log(U/\ell)])$ pieces of
polynomial bit length and subgradients within $\mu$ of $\log t$.
For the update center $u=h\odot r^{-\theta}$, we approximate each
$-\log u_e$ by a rational $c_e$ within $\mu$ and solve the rational LP
\[
 h^+\in\argmin_{h\in\mathcal K_\psi}\sum_e\bigl[\varphi(h_e)+c_eh_e\bigr].
\]
Within a layer, these LPs share their constraint matrix, so their optimal
vertices have a uniform polynomial encoding bound. Optimality supplies a
subgradient $s$ of the objective with $\ip{s}{h-h^+}\ge0$ on
$\mathcal K_\psi$, and $\|s-\nabla F(h^+)\|_\infty\le2\mu$ for
$F(h)=\KL(h\Vert u)$. Hence
\[
 \sup_{h\in\mathcal K_\psi}\ip{\nabla F(h^+)}{h^+-h}\le4M\mu\le c\tau A.
\]
By the Bregman identity and $\|\log r\|_\infty\le\log2$, the numerical
update and the smoothing $b\mapsto b^\psi$ cost at most
$c\tau A+2\psi M\log2$ in~\eqref{piece:progress}. Since an accepted
update has $\Delta_t\ge\tau A$ on the validation-noise accuracy event,
every accepted update still decreases the potential by $\Omega(\tau A)$.

\subsection{Finite precision and privacy}\label{sec:impl-coupling}

A finite-bit sampler cannot reproduce a continuous law. We therefore
use a reference mechanism with exact continuous sampling followed by
randomized rounding. We first check how rounding affects privacy and
progress, then couple the finite-bit implementation to this reference.

For dyadic $\Delta>0$, a power of two $K$, and a fresh seed
$\upsilon$ uniform on $\{0,1/K,\ldots,(K-1)/K\}$, we use the randomized
rounding $\mathsf Q_\Delta(x,\upsilon)=\Delta\lfloor x/\Delta+\upsilon\rfloor$,
applied coordinatewise with independent seeds. It has error at most
$\Delta$ and preserves nonnegativity. If random vectors
$z,\widetilde z\in\R^d$ can be coupled so that
$\|z-\widetilde z\|_\infty\le\rho\le\Delta$ except with probability
$\eta$, and the seeds are independent of $(z,\widetilde z)$, then
\begin{equation}\label{impl:rounding-coupling}
 d_{\mathrm{TV}}\bigl(\mathsf Q_\Delta(z,\upsilon),
 \mathsf Q_\Delta(\widetilde z,\upsilon)\bigr)
 \le\eta+d\Bigl(\frac\rho\Delta+\frac1K\Bigr),
\end{equation}
because, with shared seeds, a coordinate can round differently only if
its seed falls in an interval of length at most $\rho/\Delta$ modulo
one. We reserve a total-variation budget $\nu$ and take $\Delta$ to be
the largest power of two satisfying
\begin{equation}\label{impl:grid}
 \Delta\le\frac{c\gamma\nu}{Q_*},
 \qquad\text{where}\quad
 0<\nu\le c\min\Bigl\{\frac\delta{1+e^\eps},\,\beta,\,1\Bigr\}.
\end{equation}
Since $Q_*\ge Q_{\mathrm c}\ge n^2$, $\tau=\Theta(\gamma)$, $A_j\ge c\sigma/\gamma$,
$\overline M=O(n^2\sigma/\gamma)$, and $\lambda_j\ge5U$, this gives
$\overline M\Delta\ll\tau A_j$, $\Delta\ll\tau\lambda_j$, and
$Q_{\mathrm{c}}\Delta\le c\nu$.

Clamping rounded degrees to $[0,(n-1)U]$ and replacing $R_j$ by
$R_j+|V_j|\Delta$ in $\mathcal K_j$ and $A_j$ restores the confidence-set
guarantees; rounding heavy-edge values changes each cut by at most
$\binom n2\Delta$.

Rounding a released value is postprocessing, but validation uses rounded
Gaussian coordinates. Their features satisfy $\overline g_e\le2g_e+2\Delta^2$ and
$\log\overline r_e\le\tau\overline g_e$. The standardized validation
shift is therefore at most $Cs(g_\pi+\Delta^2)/\lambda$, with $g_\pi$ as
in the proof of Lemma~\ref{privacy:trial}, and since
$\E[(g_\pi+\Delta^2)^2e^{t(g_\pi+\Delta^2)}]\le C$ for $0\le t\le c$,
that proof retains the privacy cost $C\alpha s^2/\lambda^2$.

For progress, the maps $t\mapsto\log r(t)$ and $t\mapsto1-r(t)^{-\theta}$,
where $r(t)=1+\tau t^2/(4+\tau t^2)$, are $C\sqrt\tau$-Lipschitz, so
rounding changes a score by $O(\overline M\Delta)\ll\tau A$; a slightly
larger threshold constant absorbs this loss.

We now approximate each sampling step of the reference mechanism to
total-variation error $O(\nu/Q_*)$, given the common history.
By Lemma~\ref{oracle:lem-kkt}, the mass bounds, and
$\lambda_j\ge5U$, the oracle optimizer satisfies $cn^{-2}I\preceq
X\preceq nI$; on the relaxed range $(c/2)n^{-2}I\preceq Z\preceq nI$, the
negative objective is $\lambda/n^2$-strongly convex, so certified convex
optimization~\cite{NN94} computes a rational $\widetilde X$ with
$\|\widetilde X-X\|_F\le c\nu/(Q_*n^2)$, which changes the Gaussian law by
$O(\nu/Q_*)$ by Pinsker's inequality.

We sample $N(0,\widetilde X)$ by an
approximate Cholesky factor and Box--Muller transforms, and Laplace noise
by its inverse CDF, using midpoint seeds with
$O(\log(nQ_*/(\nu\Delta))+\log(1+\text{scale}))$ random bits per uniform
and certified arithmetic; midpoint seeds keep every sample bounded by a
polynomial on every execution. Outside endpoint
intervals of probability $\eta=c\nu/Q_*$, these maps have polynomially
bounded derivatives, which yields coordinate error $\rho=\eta\Delta/d$,
and~\eqref{impl:rounding-coupling} with $K\ge d/\eta$ bounds the error
of each rounded sample by $3\eta$. The validation center involves
logarithms, which we compute to error $c\nu\tau\lambda/Q_*$; this
changes the Laplace law of scale $c_{\mathrm{v}}\tau\lambda$ by
$O(\nu/Q_*)$.

The final selection of each layer
samples from scores computed to accuracy $c\nu/Q_*$ in log weight,
through cumulative probabilities floored to a dyadic grid of step
$O(\nu/((N+1)Q_*))$. The AboveThreshold instances need a separate
argument, because their hidden threshold noise persists across queries:
under a coupling of the continuous and rounded threshold noises within
$\Delta$, each comparison differs with probability $O(\eps_0\Delta)$,
since the fresh query noise has density $O(\eps_0)$.

Summing over at most $Q_*$ calls and using~\eqref{impl:grid}, an
adaptive coupling places the output law of the implementation within
total variation $\nu$ of the reference law. Consequently, if the
reference mechanism is $(\eps,\delta_0)$-DP, the implementation is
$(\eps,\delta_0+(1+e^\eps)\nu)$-DP, and its utility failure probability
is larger by at most $\nu$. The choice of $\nu$ in~\eqref{impl:grid}
fits within the budgets left by Proposition~\ref{global:privacy}, and
all operations have polynomial bit complexity, as required by
Theorem~\ref{thm:main}.

\end{document}